\documentclass[aps,showpacs,onecolumn,superscriptaddress,groupedaddress]{revtex4-1}  % for review and submission
\usepackage{graphicx}  % needed for figures
\usepackage{dcolumn}   % needed for some tables
\usepackage{bm}        % for math
\usepackage{amssymb}   % for math
\usepackage{amsmath}   % for math
\usepackage{natbib}
\usepackage{color}

\begin{document}

%\title{Active stress generation in a rearranging filament network}
\title{Role of turn-over in active stress generation in a filament network}
%\title{Active stress generation in a filament network with turnover}
\author{Tetsuya Hiraiwa$^{1,2,3}$,Guillaume Salbreux$^{1,4}$}
%\email[]{hiraiwa@zedat.fu-berlin.de}
\address{$^1$ Max Planck Institute for the Physics of Complex Systems, Dresden, 01187, Germany}
\address{$^2$ Fachbereich Physik, Freie Universit\"{a}t Berlin, Berlin, 14195, Germany}
\address{$^3$ Department of Physics, Graduate School of Science, The University of Tokyo, Tokyo, 113-0033, Japan}
\address{$^4$ The Francis Crick Institute, 44 Lincolns Inn Fields, London, WC2A 3LY, United Kingdom}
%\address{$^3$ Department of Physics, Graduate School of Science, The University of Tokyo, Hongo, Tokyo 113-0033, Japan}
%\email[]{salbreux@pks.mpg.de}
%\affiliation{Max Planck Institute for the Physics of Complex Systems, Dresden, 01187, Germany}

\date{\today}

\begin{abstract}
We study the effect of turnover of cross linkers, motors and filaments on the generation of a contractile stress in a network of filaments connected by passive crosslinkers and subjected to the forces exerted by molecular motors. We perform numerical simulations where filaments are treated as rigid rods and molecular motors move fast compared to the timescale of exchange of crosslinkers. We show that molecular motors create a contractile stress above a critical number of crosslinkers. When passive crosslinkers are allowed to turn over, the stress exerted by the network vanishes, due to the formation of clusters. When both filaments and passive crosslinkers turn over, clustering is prevented and the network reaches a dynamic contractile steady-state. A maximum stress is reached for an optimum ratio of the filament and crosslinker turnover rates.
\end{abstract}

\maketitle

%%%%%%%%%%%%%%%%%%%%%%%%%%%%
%        Introduction      %
%%%%%%%%%%%%%%%%%%%%%%%%%%%%

%%% Backgournd --- Cortical cytoskeleton
The cell cortical cytoskeleton is essential in processes involving cell shape changes \cite{Guillaume12,Joanny09}.
In the cortex, myosin molecular motors are assembled in bipolar filamentous structure which bind to actin filaments and generate forces by consuming the chemical energy of the hydrolysis of adenosine triphosphate (ATP). The action of myosin motors result in the generation of an active, contractile stress, whose spatial distribution in the cortex plays a key role in cellular morphogenetic processes \cite{mayer2010anisotropies, behrndt2012forces}. 

In living cells, passive, active crosslinkers and actin filaments are continuously exchanged between the cortex and the cytosol \cite{Guillaume12}.  As a result, cytoskeletal networks can release elastic stresses stored in the network and undergo large-scale flows. Significant progress has been achieved trough {\it in vitro} studies and theoretical analysis of actomyosin networks to understand stress generation in networks with permanent filaments and fixed or unbinding crosslinkers \cite{liverpool2009mechanical, e2011active, Dasanayake11, Lenz12,Lenz12-2, Koehler12, alvarado2013molecular}.
It is unclear however what is the role of turnover in stress generation and how filament networks can simultaneously rearrange and exert a permanent internal active stress.  {\it In vivo} experiments suggest that the rate of turn-over is a major determinant of force generation by actomyosin networks \cite{Guha:2005aa, Tinevez:2009aa}. 

\begin{figure}[h!]
 \centering
 \includegraphics[width=9cm]{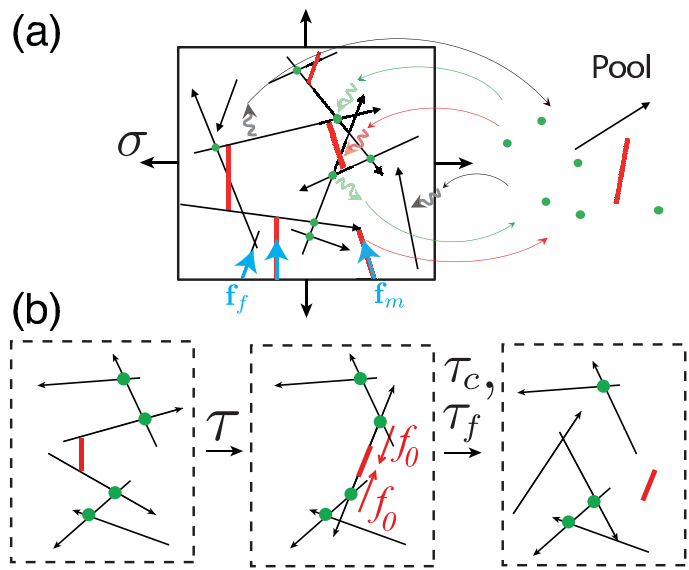}
 \caption{(a) Schematic illustration of a 2D cytoskeletal network (left).
Filaments are represented by black arrows, motors by red bars and cross linkers by green dots. The network exerts a stress $\sigma$ on the boundaries of the box, arising from tensions acting within the filaments ($\mathbf{f}_f$) and within the motors ($\mathbf{f}_m$).
Motors move towards the arrowhead of filaments (for actin, the arrowhead corresponds to the barbed end).
Network components stochastically exchange with a reservoir (right).
(b) Motors move on filaments and rearrange the network on a timescale $\tau$. Crosslinkers and actin filaments turn over on longer timescales $\tau_c>\tau$ and $\tau_f>\tau$.}
 \label{fig:schematic}
\end{figure}

We ask here how the rate of turn-over of passive crosslinkers and actin filaments influence the active stress generated by motors in the network. 
We study a simplified mechanical model for a cytoskeletal network in two dimensions 
whose constituents are turning over (Fig. \ref{fig:schematic}a).
Filaments (actin filaments) and motors (myosin motors) are treated as rigid rods. 
Filaments are assumed to have a polarity represented by the arrows as shown in Fig. \ref{fig:schematic}.
Passive crosslinkers are assumed to be point-like and constrain the position of the filaments on which they are attached. Filaments are able to rotate freely around the cross linker position and around motor heads. 
Motor heads exert an active force ${\bm f_m}$ with a constant magnitude $|{\bm f_m}|=f_0$ on filaments, 
oriented toward the reverse direction of the arrow (the minus end of actin filaments). 
%How the active stress depends on the properties of the microscopic components of the network is however not clear.  Significant progress has been made recently in understanding why this stress is generally positive, despite the fact that motors can in principle exert expansile as well as contractile force dipoles. Non-linear elastic properties of filaments have been proposed to be at the origin of the network contractile behaviour, as networks of semi-flexible filaments can sustain positive stresses but not compressive forces \cite{liverpool2009mechanical, e2011active, Lenz12,Lenz12-2}. Networks such as the actin cortex are however made of short filaments, which are likely to behave as rigid rods. It was shown in Ref. \cite{Dasanayake11} that 
%a {\it permanently} crosslinked isotropic network of rigid filaments can exhibit contractility due 
%instabilities of expansile configurations for myosins. 

%We also ask how the stress generated by a network of filament depends on the concentration of crosslinkers, which appears to play a major role in reconstituted actin networks \cite{Bendix08,Koehler12}
%Also, it is known that dependecy of contractility on the crosslinker concentration in a reconstituted actin cytoskeleton \cite{Bendix08,Koehler12}.
%Here it is unclear 
%(i) whether motors induce contractile stress when the amount of passive crosslinkers are limited in an isotropic network, 
%and (ii) how microscopic properties of the actin network can give rise to a fluid behaviour and simultaneously exert a permanent internal active stress.

To obtain forces acting on filaments, we introduce the effective mechanical potential,
$U= W + {\bm \lambda} \cdot {\bm g}({\bm x}_{f,i},{\bm n}_{f,i},{\bm x}_{m,k},{\bm n}_{m,k})$, where
 $W=-f_0 \sum_{\langle k,i \rangle} s_{ki}$ is the work due to motor active forces, 
with $s_{ki}$ the position of the $k$-th motor head on the $i$-th filament relative to the filament centre of mass, and the
sum $\sum_{\langle k,i \rangle}$ is performed for all the pairs of filaments ($i$) and motors ($k$) connected with each other. 
In addition, geometrical constraints arise from the conditions that cross linkers and motor heads are firmly attached to filaments, and that motor filaments have a fixed length. The coefficients ${\bm \lambda}$ are Lagrange multipliers imposing these constraints, ${\bm g}({\bm x}_{f,i},{\bm n}_{f,i},{\bm x}_{m,k},{\bm n}_{m,k})={\bm 0}$,
where ${\bm x}_{f,i}$ and  ${\bm n}_{f,i}$ (resp. ${\bm x}_{m,k}$, ${\bm n}_{m,k}$) indicate the centre of mass position and orientation unit vector of the $i$-th filament (resp. $k$-th motor). The motion of motors is taken into account by writing that the position of attachment of the $k$-th motor  $s_{ki}$ relative to the centre of mass of the filament $i$ follows the dynamic equation
\begin{equation}
\label{DynamicEquation}
\mu \frac{ds_{ki}}{dt}=- \frac{\partial U}{\partial s_{ki}}
\end{equation} 
with $\mu$ a scalar friction coefficient arising from translational friction between the motor heads and the filament \cite{Tawada91,Imafuku99,Julicher:1995mb}. Motors have a typical velocity $v_m=f_0/\mu$, 
and we introduce a reference timescale $\tau=l_f  \mu/(2 f_0)$. We neglect viscous forces arising from the fluid around the network compared to motor-filament friction, and the position and orientation of filaments $\mathbf{x}_{f,i}$, $\mathbf{n}_{f,i}$ are relaxed instantaneously.

To fix ideas, cortical actomyosin networks in a cell have a mesh size  $\xi \sim 20-250$nm  \cite{Morone:2006uq,Charras:2006yq}
and the typical cell diameter is several tens of micrometers. 
We therefore expect actin filaments to have a length $l_f$ of order $0.1-1\mu$m, 
smaller than their persistence length $l_p=16\mu$m \cite{ott1993measurement}. 
Myosins move on actin filaments with velocity $v_m\sim 0.1-3 \mu$m/s \cite{ishikawa2003polarized, kubalek1992dictyostelium}. 
The characteristic time for the myosins to move on a filament is $\tau \sim l_f/v_m\sim 0.03-10$s. 
To compare this timescale to the effect of viscous stresses arising in the solvent of viscosity $\eta$, 
we note that the velocity of a filament 
in the network subjected to a force $f_0$ is $\sim f_0 \ln (\xi/r_f)/ (\eta l_f) $, 
giving a timescale $\tau_v \sim l_f^2 \eta/(f_0 \ln (\xi/r_f))$, with $r_f\simeq 5nm$ the radius of a filament. 
Taking the force exerted by a myosin $f_0\sim6-12 pN$, $\eta\simeq 10^{-3} Pa.s$ for water, 
we find $\tau_v \sim 10^{-4}$s $\ll\tau$. Finally, we expect the characteristic times $\tau_c$ and $\tau_f$ of the crosslinker and actin filament turnover, respectively, to be of the order of $10$ s- $1$ min \cite{Mukhina:2007ve, Reichl:2008fr}, slow compared to these timescales, such that $\tau_c \gg \tau$ and $\tau_f \gg \tau$.

%%%%%%%%%%%%%%%%%%%%%%%%%%%%%%%%%%%%%
%                                      Microscopic theory                                            %
%%%%%%%%%%%%%%%%%%%%%%%%%%%%%%%%%%%%%

\begin{figure}[h!]
 \centering
 \includegraphics[width=11cm]{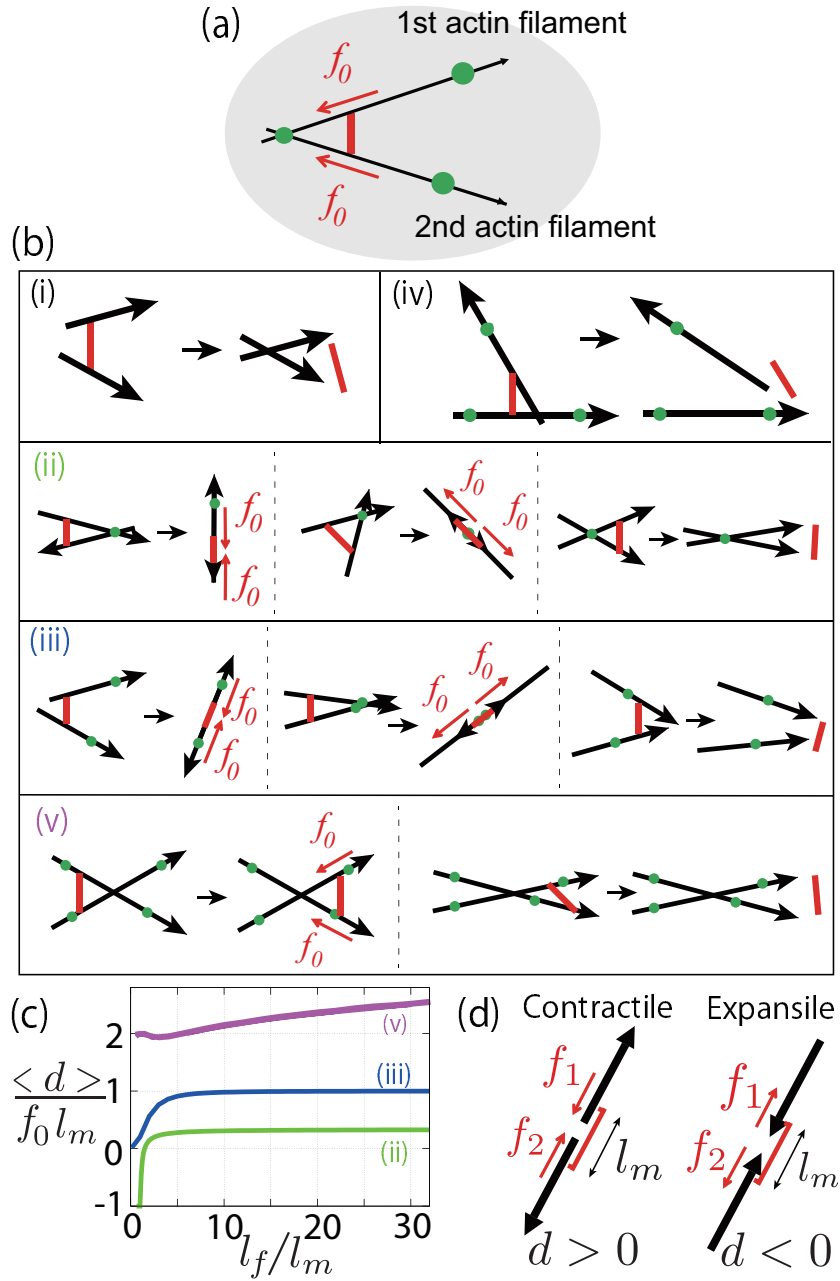}
 \caption{Contractile and expansile configurations for one motor and two filaments.
(a) Two filaments are attached by fixed cross linkers to the external network.
(b) (Left) Possible initial configurations for one motor, two filaments and one or several cross linkers, and final configurations.
(c) Average strength of the contractile force dipole generated by the molecular motor $\langle d \rangle$ in the final configurations (ii), (iii), (v) where the motor does not detach,
as a function of the filament length $l_f$.
(d) Contractile (force dipole $d>0$) and expansile ($d<0$) configurations.
%(e) Fraction $\phi$ of initial conditions relaxing to a steady state where the myosin stats attached to the filaments.
}
 \label{fig:theory}
\end{figure}

%How the active stress depends on the properties of the microscopic components of the network is however not clear.  Significant progress has been made recently in understanding why this stress is generally positive, despite the fact that motors can in principle exert expansile as well as contractile force dipoles. Non-linear elastic properties of filaments have been proposed to be at the origin of the network contractile behaviour, as networks of semi-flexible filaments can sustain positive stresses but not compressive forces \cite{liverpool2009mechanical, e2011active, Lenz12,Lenz12-2}. Networks such as the actin cortex are however made of short filaments, which are likely to behave as rigid rods. It was shown in Ref. \cite{Dasanayake11} that 
%a {\it permanently} crosslinked isotropic network of rigid filaments can exhibit contractility due 
%instabilities of expansile configurations for myosins. 

\paragraph{Configuration with one motor and two filaments.} 
% It was shown in Ref. \cite{Dasanayake11} that 
%a permanently crosslinked isotropic network of rigid filaments can exhibit contractility due to
%instabilities of expansile configurations for myosins.
We start by discussing the dynamics of configurations involving
one motor attached to two filaments (Fig. \ref{fig:theory}a). The two filaments can be connected to a crosslinker, itself connected to the external network, which we consider here to be fixed in space.
%we smooth out the network components other than 
%focused one myosin mini-filament (red bar), associated two filaments (black arrows), 
%and passive crosslinkers (green points) which connect the associated filaments with the background network or each other,
%as illustrated in Fig. \ref{fig:theory}(a).
%We suppose that crosslinkers on the background network is fixed in the space.
%We further assume that turnover of passive crosslinkers are much slower than the motor motion, {\it i.e.} $\tau_c \gg \tau$. In that limit, motors relax quickly compared to cross linker exchanges, and a fraction $\frac{\tau_c}{\tau+\tau_c}\simeq1$ of motors are in a steady state configuration. 
We distinguish several basic possible configurations, depending on 
the number and positions of attached passive crosslinkers (Fig. \ref{fig:theory}b):
both filaments can be completely free (i), the two filaments can be crosslinked to each other (ii), or they can be crosslinked to the external network (iii)-(v).
%We name the motives in Fig. \ref{fig:theory}(b)-(f) the free motif, one-hinge motif, two-hinge motif, half-rigid motif, and rigid motif, respectively, from (b) to (f).

To investigate how molecular motors modify the filaments organisation, we study the relaxation to final state of these different configurations. By averaging over possible initial configurations, we evaluate whether motors form on average positive or negative force dipoles in the network (Fig. \ref{fig:theory}c).  
The motor force dipole is $d=- (l_m/2) \mathbf{n} \cdot (\mathbf{f}_1-\mathbf{f}_2)$, with $\mathbf{n}$ the unit vector giving the motor orientation, and $\mathbf{f}_1$ and $\mathbf{f}_2$ are 
the forces by which the motors pull or push the filaments (Fig. \ref{fig:theory}d).
In case (v), $\mathbf{f}_1$ and $\mathbf{f}_2$ include not only the forces exerted by the motor themselves
but also the forces originating from the geometrical constraint (Appendix \ref{AverageForceDipole}).
%To compute the average force dipole created by the motor in each configuration, we 
%average over uniform distributions for the initial positions of the centers of mass of filaments,
%initial angles of two filaments ${\bm \theta}_{a,i}$,
%initial position of the motor on the filaments $s_i$,
%and attached positions of the passive crosslinkers.
We find that possible initial configurations can be classified in 2 categories (Fig. \ref{fig:theory}b). In cases (i) and (iv), filaments are moved relative to each other until the motor detaches, so that a motor-induced force dipole acts on a transient time $\tau$. This force dipole is expansile on average. In the quasi-static limit where $\tau \ll \tau_c$, the contribution of transient filament-motor configuration to the overall stress is negligible compared to steady motor configurations. When filaments either (ii) have only one fixed attachment point to the external network, (iii) are connected to each other by a cross linker, or (v) have both more than 2 cross linkers, they relax to a steady configuration where the motor exerts a constant force dipole. By averaging the resulting force  dipole over possible initial configurations of the two filaments and the motor (Appendix \ref{AverageForceDipole}), we find that the motor exerts a contractile force dipole on average in cases (iii) and (v) (Fig. \ref{fig:theory}c). As in Ref. \cite{Dasanayake11}, the bias towards contractile states arises from instabilities of expansile configurations (Appendix \ref{AverageForceDipole}). 
The average force dipole $\langle d\rangle$ vanishes for point-like motors, $l_m\rightarrow 0$ \cite{lenz2014geometrical}.

%As consequences, we obtain the filaments length $L_a(=L_1=L_2)$-dependency of the magnitude of the isotropic force dipole, $d \equiv (d_{xx}+d_{yy})/2$,
%as shown in Fig. \ref{fig:theory}(h).
%We skip the motif shown by \ref{fig:theory}(e) because, when the system has passive crosslinkers, 
%the motif (e) has no stable steady configurations so that only the motives (b), (c) or (f) can be realized at the final states.
%The motives (b) and (c) yield extensile force dipoles (solid line and triangles, respectively, in Fig. \ref{fig:theory}(h)),
%whereas the motives (d) and (f) yield contractile force dipole (crossmarks and circles, respectively, in Fig. \ref{fig:theory}(h)).
%The stable point in rigid motif is consistent with that in Ref. \cite{Dasanayake11}.
%
% In motives shown by Figs. \ref{fig:theory}(c), (d) and (f), 
%the system can stay at the stable or neutral fixed states.
%Therefore, we average $d_{kl}$ at those steady states
%uniformly with respect to attached positions of the passive crosslinkers,
%initial positions of the centers of mass of F-actins $\Delta {\bm c}_1(t=0)$,
%initial angles of two filaments ${\bm \theta}_{a,i}(t=0)$,
%and all possible initial myosin-head-attached positions $s_i(t=0)$.
%On the other hand, the system cannot stay at any stable or neutral fixed configuration for the motives in Figs. \ref{fig:theory}(b) and (e).
%

%%%%%%%%%%%%%%%%%%%%%%%%%%%%%%%%
%                                       Simulation                                        %
%%%%%%%%%%%%%%%%%%%%%%%%%%%%%%%%
\paragraph{Numerical simulations of networks with turnover.}  
%The total number of filaments, motors and passive crosslinkers, $N_f$, $N_m$ and $N_c$ is fixed.
We next numerically simulated a network of $N_f$ filaments, $N_m$ motor rods and $N_c$ passive crosslinkers
in a square box of width $W$ with periodic boundaries (Fig. \ref{fig:schematic}). The frame of the box is not allowed to deform. 
We fixed the normalised motor density $l_m^2 c$ to 1, where $c=N_m/W^2$ is the motor density. 
To make numerical simulations easier, the geometrical constraints ${\bm g}$ were replaced by linear springs mimicking the contacts at the junctions between motors and filaments, and at crosslinking points. 
%A small friction acting on the position and orientations of filaments was added to simulate quasi-static relaxation of the network. 
Initial conditions are obtained by randomly positioning filaments in the network, and timescales are normalized to the reference time $\tau$. 
Turnover is introduced by stochastically removing cross linkers, motors and filaments from the network with rates $1/\tau_c$, $1/\tau_m$ and $1/\tau_f$. 
For simplicity, turnover rates of crosslinkers are taken here independent of the forces they sustain \cite{alvarado2013molecular}. Filaments, motors and passive crosslinkers are added in the network with on-rates $k_{on}^f$, $k_{on}^m$ and $k_{on}^c$. New filaments take random positions and random orientation, motors take a randomly chosen position on two filaments points separated by a distance $l_m$, and passive crosslinkers are put on a randomly chosen filament intersection.

We evaluate the components of the stress tensor $\sigma_{ij}$ ($i,j=x \text{ or } y$) acting on the boundary of the box (Fig. \ref{fig:schematic}a).  The total stress is given by two contributions $\sigma_{ij}=\sigma_{ij}^f+\sigma_{ij}^m$, with $\sigma_{ij}^f$ obtained by summing forces acting both within the filaments and $\sigma_{ij}^m$ from forces acting within motor rods crossing the boundary of the box (Fig. \ref{fig:schematic}). 
In a linear elastic or viscous material, $\sigma_{ij}^f=0$ in the absence of large-scale deformation.
%, and the stress reduces to $\sigma^m_{ij}$. 
In a non-linearly elastic material however, $\sigma_{ij}^f \neq 0$ (Appendix \ref{StressAverageForceDipole}).
% and the total stress does not reduce to Eq. (\ref{StressForceDipoles}) (Supp Mat).
At steady state, the average stress acting within the motors is given by (Ref.  \cite{aditi2002hydrodynamic} and Appedix \ref{StressAverageForceDipole})
\begin{eqnarray}
\sigma^m_{ij}&\simeq& \langle d   n_{i}n_{j} c \rangle \ ,
\label{StressForceDipoles}
\end{eqnarray}
where $\mathbf{n}$, $d$ and $c$ denote the orientation, motor-induced force dipole strength and concentration of bound motors. 
%Strength of the force dipole $d$ exerted by each motor is evaluated as $d = (l_m/2) f_0 \cos \psi_1 + (l_m/2) f_0 \cos \psi_2$ with the angles $\psi_1$ and $\psi_2$ between the motor and associated two filaments.
When the dipoles are isotropically oriented, $\sigma^m_{ij}=\sigma^m\delta_{ij}$ with $\sigma^m \simeq \langle d c \rangle/2$.

\begin{figure}[t!]
 \centering
 \includegraphics[width=0.9\columnwidth]{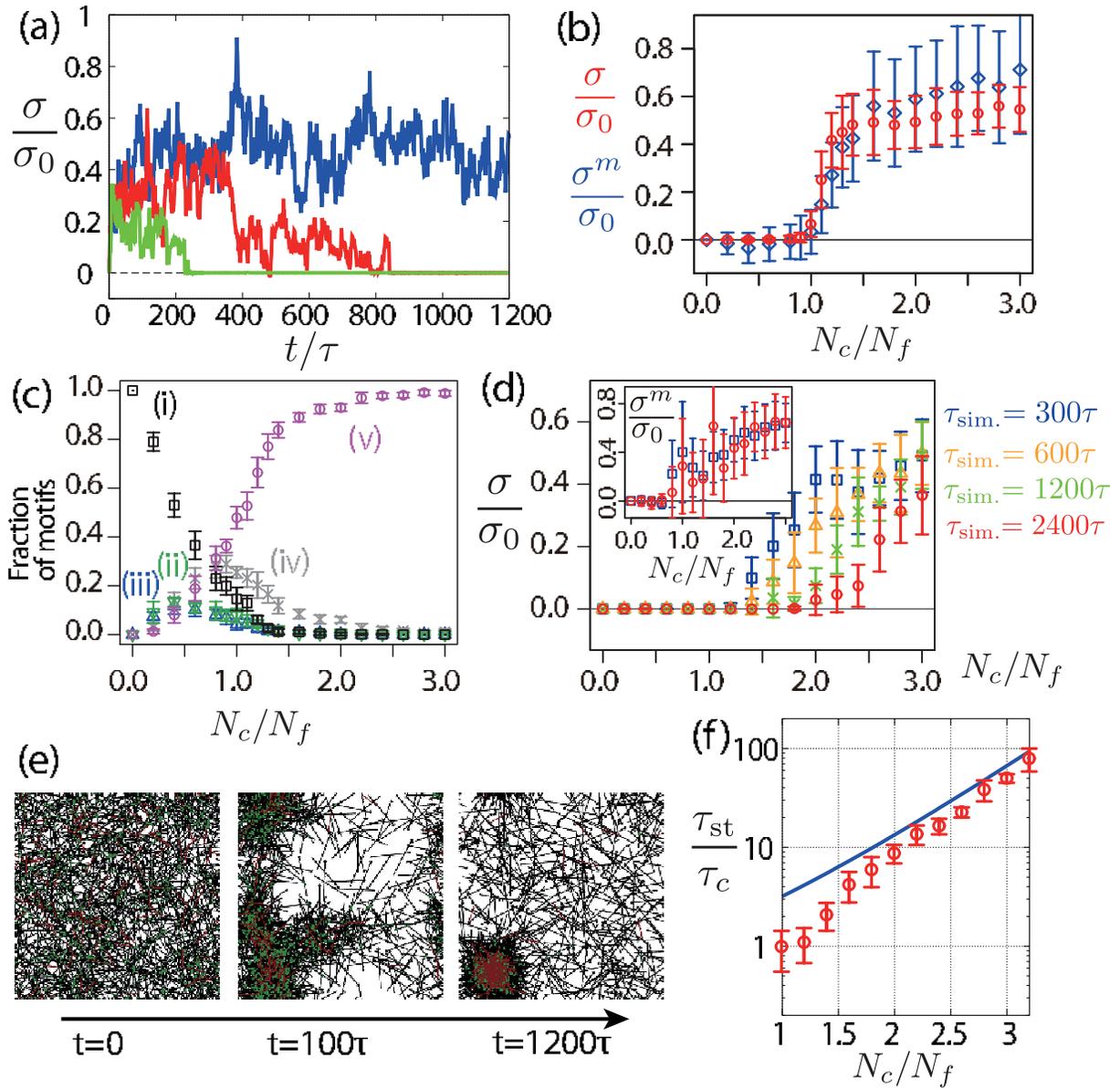}
 \caption{Stress in a network of motors and filaments, with and without crosslinker turnover.
 %(a) Schematic of numerical simulation: the network is simulated in a box of size $W$ with periodic boundary conditions. The isotropic stress $\sigma$ is obtained by measuring the normal force acting on the boundaries of the box.
(a) Time evolution of the normalised stress $\sigma/\sigma_0$ for $\tau_c^{-1}=0$, $N_c=1.2 N_f$ (blue), $\tau_c=100 \tau$, $N_c=1.6 N_f$ (red) and $\tau_c=100 \tau$, $N_c=1.2 N_f$ (green). ($\sigma_0 = f_0 l_m N_m/W^2$)
(b) Steady-state isotropic stress $\sigma$ (circles) and motor stress $\sigma^m$ (cross marks)
as a function of the crosslinkers number $N_c$ for $\tau_c^{-1}=0$.
(c) Fraction of different configurations (i)-(v) in Fig. \ref{fig:theory} at steady state as a function of the crosslinker number $N_c$ for $\tau_c^{-1}=0$.
(d) Isotropic stress $\sigma$ as a function of the crosslinker number $N_c$, for finite crosslinker lifetime $\tau_c=100\tau$, and for several simulation times ($\tau_{\rm sim.}=300, 600, 1,200$ and $2,400 \tau$). 
Stress within the motors  $\sigma_m$ is shown in the inset for $\tau_{\rm sim.}=300$ and $2,400 \tau$.
(e) Snapshots of a simulated network with crosslinker turnover ($N_c=1.2 N_f$, $\tau_c=100\tau$). 
(f) Stress decay time $\tau_{\rm st}$ as a function of the crosslinker number $N_c$ with crosslinker turnover ($\tau_c=100\tau$). 
Solid line, theoretical prediction (see main text).
%The stress decaying time is obtained by fitting the time-evolution of the stress $\sigma(t)$ by $\sigma = a \exp(-t/\tau_{\rm st.}) +b$ with fitting parameters $a$, $b$ and $\tau_{\rm st.}$.
Other parameters: $N_f=1,000$, $N_m=100$, $\tau_{f}^{-1}=0$, $k_{on}^c\tau_c=20$, $l_f=2l_m$, $W=10 l_m$, $\tau_m=100 \tau$ and $k_{on}^m\tau=20$. }
 \label{fig:numwithoutAT}
\end{figure}

We first performed simulations without turnover of filaments. We focus on the isotropic component of the stress $\sigma \equiv (\sigma_{xx}+\sigma_{yy})/2$. 
Figure \ref{fig:numwithoutAT}a (blue curve) shows the typical time evolution of the stress, for permanent crosslinkers and for crosslinkers with a finite lifetime.
Without crosslinker turnover, the system reaches a stationary state where the stress $\sigma$ fluctuates around a finite positive value. 
Figure \ref{fig:numwithoutAT}b shows the average value of the resulting stress (circles) as a function of the number of cross linkers $N_c$.
For a small number of crosslinkers ($N_c \rightarrow 0$), no stress is observed, $\sigma=0$. 
Above a critical value of the number of crosslinkers $N_c>N_c^*\simeq N_f$, a transition occurs and a positive contractile stress appears in the system. 
When $N_c \sim N_c^{*}$ but $N_c<N_c^{*}$,
although a portion of motors generate contractile force dipoles, 
filaments aggregate in clusters and hence $\sigma \sim 0$, 
suggesting that the transition is associated to the network connectivity (Supp movies M1 and M2). 
The stress then further increases with the number of crosslinkers and eventually saturates to a positive value. Such a transition to contractility as a function of the number of crosslinkers has been reported in {\it in vitro} reconstituted networks \cite{Bendix08, Koehler12}, as well as network clustering \cite{kohler2011structure}.

The average stress only within the motors $\sigma^m$ is also plotted in Fig. \ref{fig:numwithoutAT}b (cross marks). 
A similar transition occurs for a critical value of the number of cross linkers. 
The stress $\sigma^m$ is however larger than the total stress for a large number of cross linker $N_c>N_f$, 
implying that the filament network is under compression ($\sigma^f< 0$). 
For large $N_c$, configurations (v) dominate in the network (Fig. \ref{fig:numwithoutAT}c). 
The saturating value of $\sigma^m/\sigma_0 \sim 0.6$ is nevertheless smaller than the average force dipole obtained by averaging all possible filament orientations $\langle d/2 \rangle / (f_0 l_m) \sim 1.0$ (Fig. \ref{fig:theory}c); 
this is because some configurations do not relax to equilibrium on the characteristic timescale of myosin turnover.
%The saturating value of $\sigma^m$ in Fig. \ref{fig:numwithoutAT}c is in accordance with the fact that the configuration (v) dominates for large $N_c$ (Fig. \ref{fig:numwithoutAT}c) and yields the average force dipole $\langle d/2 \rangle \sim 0.324$ (Fig. \ref{fig:theory}d) for $l_f=2 l_m$.
%Note that the total stress $\sigma$ is distinct from the average motor-induced stress $\sigma^m$, due to  nonlinear elasticity of the filament network.
%To determine the minimal number of crosslinkers generating a contractile stress and the saturating value, we perform the following calculation: (TO ADD).

%%%%%%%%%%%%%%%%%%%%%%%%%%%%%%%%%%%%%%%%%%%%%%%%%%%%
When crosslinkers are allowed to turn over, the average stress first reaches a positive value before decaying to zero (Fig. \ref{fig:numwithoutAT}a, red and green curves),
even though the stress within the motors $\sigma^m$ is still non zero (Fig. \ref{fig:numwithoutAT}d). 
%\footnote{In Fig. \ref{fig:numwithoutAT}d (inset), motor-exerted force dipole is also decreasing much slower than the stress decay.
%This suggests the gradual optimization of directionality of filaments in the cluster to weaken the contractile dipole strength.
%However, investigation on the intra-cluster structure may be beyond the capacity of this model since no direct interactions between filaments like excluded volume effect are included.}
The decay of the stress correlates with the collapse of the network in an isolated cluster, where filaments accumulate (Fig. \ref{fig:numwithoutAT}e, Supp movie M3). 
To evaluate the decay timescale $\tau_{st}$, we fitted an exponentially decreasing function $\sigma(t) = A \exp(-t/\tau_{st}) $ to the simulation results with fitting parameters $A$ and $\tau_{st}$.
The decay timescale of the stress increases exponentially with $N_c$ for large $N_c$ (Fig. \ref{fig:numwithoutAT}f). The relaxation timescale of the stress can be understood as follows: the timescale $\tau_{st}$ corresponds to the relaxation Maxwell time on which the network becomes fluid, due to turnover of passive crosslinkers enabling network rearrangements. 
This time can be estimated by
\begin{equation} \label{eq:taust}
\tau_{st}=\frac{\tau_c}{n_c^*} (e^{n_c^* }-1)
\end{equation}
with $n_c^*=2N_c/N_f$. To obtain Eq. (\ref{eq:taust}), we assume that filaments with at least one crosslinker are fixed, 
and only filaments with no crosslinker attachment can rearrange. 
We then compute the first passage time at which a filament is free from cross linkers, 
starting from a configuration where the filament is attached with only one crosslinker (Appendix \ref{App:FirstPassageTime}). 
Equation (\ref{eq:taust}) accounts for the characteristic time of stress decay in the network for large values of $N_c/N_f$ (Fig. \ref{fig:numwithoutAT}f).
%See Supplementary Material for the derivation of Eq. (\ref{eq:taust}).
%%%%%%%%%%%%%

%%%%%%%%%%%%%%%%%%%%%%%%%%%%%%%%%%%%%%%%%%%%%%%%%%%%%%%%%%%%%%%
% actin turnover
\begin{figure}[t!]
 \centering
 \includegraphics[width=0.9\columnwidth]{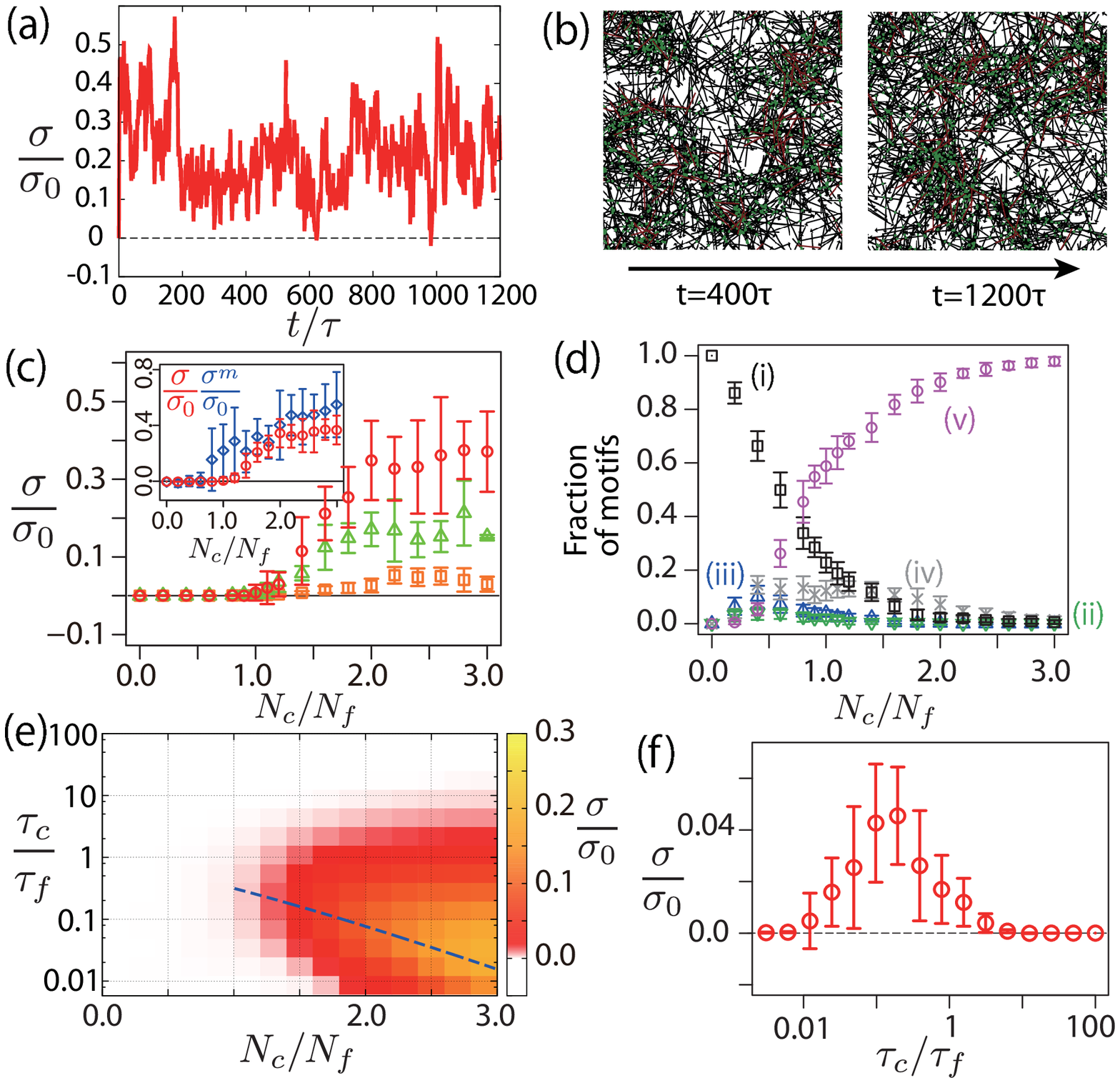}
 \caption{Stress in a network with filament turnover.
(a) Time evolution of the stress in a network with crosslinkers and filament turning-over
for $\tau_f=100\tau$, $\tau_c=100\tau$ and $N_c/N_f=1.6$.
(b) Snapshots of the network evolution corresponding to (a).
(c) Steady-state stress $\sigma$ as a function of the number of crosslinkers for $\tau_c=\tau_f=100 \tau$ (circles), $\tau_c=\tau_f=10 \tau$ (triangles) and $\tau_c=\tau_f=1 \tau$ (squares).
Inset: steady-state stress $\sigma$ (circles) and motor stress $\sigma^m$ (cross marks) for $\tau_c=\tau_f=100 \tau$.
(d) Fraction of different configurations (i)-(v) at steady state, as a function of $N_c$ ($\tau_c=100 \tau$, $\tau_f=100 \tau$).
(e) Heat map for contractile stress $\sigma$ as a function of $N_c$ and $\tau_{c}/\tau_{f}$ ($\tau_c=\tau$).
The broken line indicates $\tau_f=\tau_{\rm st}$.
(f) Stress as a function of the ratio of filament and crosslinker turnover rate ($\tau_c=\tau$, $N_c=1.6 N_f$).
 Other parameters as in Fig. \ref{fig:numwithoutAT}, except $\tau_m^{-1}=0$.}
 \label{fig:numwithAT}
\end{figure}

%Therefore, networks with crosslinkers which turnover can only exert significant stresses transiently. 
We now turn to simulations where both filaments and crosslinkers turn over. In the cell, actin filaments polymerise and depolymerise.  
Here, we account for this process by simply introducing a rate of filament turnover, $\tau_f^{-1}$. Passive crosslinkers and motor heads are removed together with the filaments to which they are attached. 
Remarkably, with both crosslinker and filament turnover, the network evolves towards a steady state with a non-zero positive stress (Fig. \ref{fig:numwithAT}a-b). As in the previous case, a transition from a non-contractile to a contractile network appears when the number of crosslinkers $N_c$ is increased (Fig. \ref{fig:numwithAT}c,  Supp movies M4 and M5). 
%As for the case without filament turnover, (.
Note that the stress $\sigma$ deviates again from the average stress within the motors $\sigma^m$ (Fig. \ref{fig:numwithAT}c, inset). 
The dependency of the fraction of configurations as a function of $N_c$ is qualitatively similar to the case without filament turnover (Fig. \ref{fig:numwithAT}d).

We then varied the filament turnover timescale $\tau_f$ (Fig. \ref{fig:numwithAT}c,e,f).  %When the crosslinker turnover time scale $\tau_c$ is set as same as $\tau_f$, only the amplitude of stress depends on the time scales (Fig. \ref{fig:numwithAT}c).
%On the other hand, $N_c$-dependecy is observed when we varied the ratio of actin and crosslinker turnover time scales $\tau_c/\tau_f$, with keeping $\tau_c=1 \tau$ .
We find that the stress reaches a maximum for intermediate values of the filament turnover timescale (broken line in  Fig. \ref{fig:numwithAT}e, and Fig. \ref{fig:numwithAT}f): 
for slow filament turnover $\tau_f \gg \tau_{\rm st}$, the network collapses in clusters due to crosslinker turnover, 
while for fast filament turnover $\tau_f \ll \tau_{\rm st}$, filaments are removed before motors reach a configuration where they can exert a force dipole in the network. We find that the optimum value of the ratio of turnover time scale in Fig. 4f, $\tau_f/\tau_c\sim10$, is of the order of the ratio experimentally measured values of turn-over of actin and crosslinker in the cell cortex, $\tau_f\sim 15-45$s and $\tau_c \sim7-14 $s \cite{Guillaume12}.
%For intermediate crosslinker turnover when a stress in generated in the network, transient formation of clusters and strong fluctuations in actin density are visible in simulations (Fig. \ref{fig:numwithAT}b).

%Real living cells might be taking advantage of this stabilization due to the actin turnover.
%A more realistic model of actin turnover would however have to take into account the gradual dissociation of monomers and nucleation and growth of actin filaments.

%%%%%%%%%%%%%%%%%%%%%%%%%%%%%%%
%                                   Conclusion                                    %
%%%%%%%%%%%%%%%%%%%%%%%%%%%%%%%

%From the consideration based on one motor and two filaments,
\paragraph{Conclusion.}
%This result requires us to take care ... when we assume contractility of myosin motor {\it ab initio} in the model.
We propose the following dynamic picture for the stress generated in a rearranging network: 
molecular motors move on filaments on a timescale $\tau$, fast compared to the crosslinker turnover $\tau_c$. 
When molecular motors bind to pairs of filaments, they either displace them and detach, or find steady configurations where they generate predominantly contractile force dipoles (Figs. \ref{fig:theory} and \ref{fig:numwithoutAT}a-c). 
In networks with permanent crosslinkers, a transition to contractile state of the network
occurs for a large enough number of crosslinkers (Fig. \ref{fig:numwithoutAT}b).
%When crossOn long-time scales, larger than $\tau_c$, turnover of crosslinkers and of the filaments allow the network to rearrange and to flow. 
When crosslinkers are allowed to turnover, the network can rearrange and flow,
filaments collapse in a cluster, and the total stress in the network vanishes
even though the stress only within the motors is still contractile (Fig. \ref{fig:numwithoutAT}a,d-f). 
Introducing filament turnover occurring on a timescale $\tau_f$ comparable to the crosslinker turnover timescale $\tau_c$ prevents this clustering mechanism, and allows the network to generate a steady-state contractile stress (Fig. \ref{fig:numwithAT}). It will be interesting to investigate stress generation in {\it in vitro} experiments where crosslinkers and filaments are allowed to turn over.

{\bf Acknowledgements:} We thank Matthew Smith for helpful discussions.
This work was supported partly by 
the JSPS Institutional Program for Young Researcher Overseas Visits (T.H.), 
the Postdoctoral Research Fellowship of the Alexander von Humboldt foundation (T.H.),
the JSPS Core-to-Core Program (T.H.), 
the Max Planck Gesellschaft (G.S.), 
and the Francis Crick Institute which receives its core funding from Cancer Research UK, the UK Medical Research Council, and the Wellcome Trust (G.S).

\appendix

\section{Calculation of the average force dipole for two filaments}
\label{AverageForceDipole}
\begin{figure}[h]
 \centering
 \includegraphics[width=14cm]{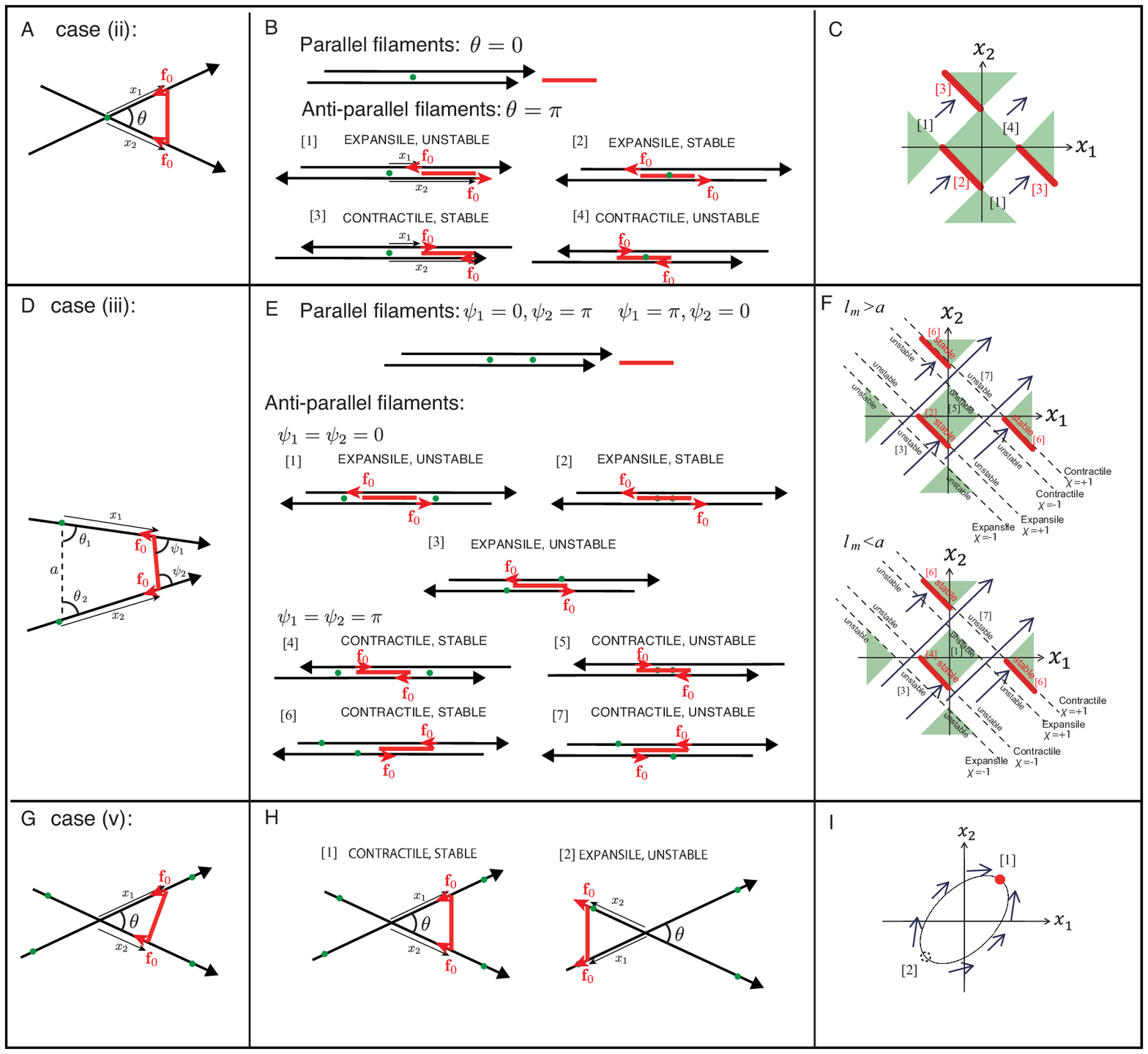}
 \caption{Final configurations reached by two filaments bound by a motor, with the filaments attached to a rigid external network with a varying number of cross linkers. A, D, G, example configurations for each case, according to the number of cross linkers; B, E, H, corresponding possible steady-state configurations; C,F,I, Phase plot of the dynamics of the motor according to the coordinates of its position on the two filaments, $x_1$, $x_2$. Red lines and dots correspond to stable steady-state configurations, while dotted black lines correspond to unstable steady-state configurations. Green regions represent values of the coordinates $x_1$ and $x_2$ which are not geometrically accessible.
 \label{fig:twofilaments}}
\end{figure}

%\paragraph{Force dipole in the configuration shown by Fig. \ref{fig:theory}(i):}
We detail here how we obtain the average force dipole exerted by a motor on two filaments in the configurations shown in Fig. 2. As mentioned in the text, we assume here that crosslinkers are rigidly fixed to the external network. The geometry of the two filaments and motor can be described the position of the motors on the two filaments $s_1$, $s_2$ and by the two angles $\psi_i$ between the motor filament and  $i$-th filament ($i=1,2$). The motor positions $s_i$ evolve according to
\begin{equation}
\label{DynamicEquation}
\mu \frac{ds_{i}}{dt}=- \frac{\partial U}{\partial s_{i}},
\end{equation} 
 while the angles $\psi_i$ relax instantaneously, so that $\partial U / \partial \psi_i =0$.
 
\subsection{No crosslinkers attached, or one filament is attached by only one crosslinker and the other filament is rigidly attached to the external network }
Case (i) and (iv): In these situations, no steady-state is reached as the motor always unbinds from the two filaments.

\subsection{Two filaments attached to each other by a crosslinker}
Case (ii): When the two filaments are attached by a single crosslinker, the mechanical potential $U$ reads
 \begin{equation}
 U=-f_0(x_1+x_2)+\lambda \left(\sqrt{x_1^2+x_2^2-2 x_1 x_2 \cos\theta}-l_m \right)
 \end{equation}
 with $x_1$ and $x_2$ the distances between the myosin attachment points to the crossing point of the two filaments, taken positive in the direction of the filament towards which motors move (Fig. \ref{fig:twofilaments}A). $\theta$ is the angle between the two filaments, such that $\mathbf{n}_1.\mathbf{n}_2=\cos\theta$, with $\mathbf{n}_i$ the unit vector giving the orientation of the filament $i$. In addition, $\lambda$ is a Lagrange multiplier ensuring that the length $l_m$ of the motor is fixed. Three situations can then occur: 
 \begin{itemize}
 \item when $\theta\neq0$ and $\theta\neq \pi$ and the filaments are not parallel neither antiparallel, the angle $\theta$ is given by
 \begin{equation}
 \label{ValuecosthetacaseII}
 \cos\theta=\frac{x_1^2+x_2^2-l_m^2}{2 x_1 x_2}
 \end{equation}
 and is free to adjust as the motor position changes. The evolution of the position of the motor on the filaments is given by 
  \begin{eqnarray}
  \label{x1evolutioncaseII}
 \mu \frac{dx_1}{d t}&=&f_0\\
   \label{x2evolutioncaseII}
  \mu \frac{d x_2}{d t}&=&f_0
 \end{eqnarray}
 such that the motor moves until it detaches from the filaments, or until the filaments become parallel or antiparallel.
 \item When the two filaments are parallel and point in the same direction, $\theta=0$ and $(x_1-x_2)^2=l_m^2$. The center of mass of the motor at position $(x_1+x_2)/2$ follows the dynamic equation
 \begin{equation}
\mu \frac{d}{dt}\frac{x_1+x_2}{2}=f_0
 \end{equation}
 and the motor moves on the filament until it detaches.
 \item When the two filaments are antiparallel, $\theta=\pi$ and $(x_1+x_2)^2=l_m^2$. Two configurations are then possible, according to whether the two filaments point away or towards the center of the motor from their attachment point to the motor. These two configurations correspond respectively to $x_1+x_2=-l_m$ (expansile configuration) and $x_1+x_2=l_m$ (contractile configuration). Any position of the motor on the filament is then a steady-state solution, as long as the two motor ends each bind on the filaments.
 
 To test for the stability of these two configurations, we consider a slight change of the angle between the two filaments, $\theta=\pi+\delta \theta$, with $|\delta\theta| \ll 1$. 
We take the initial position of the motor to be $x_1=x_0+\chi l_m /2 $ and $x_2=-x_0 + \chi l_m/2$, with $\chi=-1$ for an expansile configuration and $\chi=1$ for a contractile configuration. $x_0=(x_1-x_2)/2$ is the position of the center of mass of the motor, relative to the crosslinker joining the two filaments, measured positively in the direction of the first filament. We find then the following dynamic equation for $\delta\theta$:
\begin{eqnarray}
\mu \frac{d \delta \theta}{dt}&=& \frac{d x_1}{dt} \frac{\partial \theta}{\partial x_1}+\frac{dx_2}{dt}\frac{\partial\theta}{\partial x_2}\nonumber\\
&=&f_0 \frac{8  \chi l_m}{l_m^2- 4  x_0^2} \frac{1}{\delta \theta}\label{StabilityCaseII}
 \end{eqnarray}
where we have used Eqs. (\ref{ValuecosthetacaseII}), (\ref{x1evolutioncaseII}) and (\ref{x2evolutioncaseII}) from the first to the second line. The sign of the right hand side of Eq. (\ref{StabilityCaseII}) indicates whether the angle $\theta$ increases or decreases when the two filaments are slightly rotated away from their antiparallel configuration. Therefore, the associated configuration is unstable when the sign is positive, and stable otherwise.
 We find therefore that the stability of the configuration depends on the position of the center of mass of the motor: for $|x_0|<l_m/2$ (motor near the crosslinker), the expansile configuration is stable, and the contractile configuration unstable. For $|x_0|>l_m/2$ (motor away from the cross-linker), the expansile configuration is unstable and the contractile configuration is stable.
  \end{itemize}
The flow diagram in Fig. \ref{fig:twofilaments}C case (ii) shows the corresponding dynamics in the space of the motor position $(x_1,x_2)$. Red segments indicate the stable steady states. In the green regions, no value of the angle $\theta$ allows for the motor to have position ($x_1$, $x_2$) on the two filaments.

\subsection{Two filaments, each attached by a crosslinker to the external network}
The mechanical potential reads in that case
\begin{equation}
U=-f_0(x_1+x_2)+\lambda\left[\sqrt{x_1^2+x_2^2+l_m^2+2 x_1 l_m \cos\psi_1+2 x_2 l_m \cos\psi_2+2 x_1 x_2 \cos(\psi_1+\psi_2)}-a\right] ,
\end{equation}
with $x_1$ and $x_2$ the distances between the myosin attachment point and the crosslinker position on each filament, and $a$ the distance between the two cross linkers (Fig. \ref{fig:twofilaments}D). $\lambda$ is a Lagrange multiplier imposing that the length of the motor is equal to $l_m$. In this subsection, we use for convenience the dynamics of the angle between motor and filaments $\psi_1$ and $\psi_2$ to characterise the orientation of the filaments. The dynamics in the limit where $\psi_1$ and $\psi_2$ relax quasi statically is given by
 \begin{eqnarray}
   \label{x1evolutioncaseIII}
 \mu \frac{\partial x_1}{\partial t}&=&f_0+\frac{\lambda}{a} \left(x_1 +l_m \cos\psi_1 + x_2 \cos(\psi_1+\psi_2)\right)\\
   \label{x2evolutioncaseIII}
  \mu \frac{\partial x_2}{\partial t}&=&f_0+\frac{\lambda}{a} \left(x_2 +l_m \cos\psi_2 + x_1 \cos(\psi_1+\psi_2)\right)\\
  \epsilon \mu a^2 \frac{\partial \psi_1}{\partial t}&=& - \frac{\lambda}{a} \left(x_1 l_m \sin\psi_1+x_1 x_2 \sin(\psi_1+\psi_2)\right)    
\label{psi1evolutioncaseIII} \\
  \epsilon \mu a^2 \frac{\partial \psi_2}{\partial t}&=& - \frac{\lambda}{a} \left(x_2 l_m \sin\psi_2+x_1 x_2 \sin(\psi_1+\psi_2)\right)    
\label{psi2evolutioncaseIII}
\end{eqnarray}
where $\epsilon \ll 1$ is a factor that vanishes in the quasi-static limit where the filaments rotate adiabatically. 
Solving for the Lagrange multiplier $\lambda$ 
by imposing the constraint that the length of the motor is equal to $l_m$, we obtain:
\begin{eqnarray}
\label{LambdaEquation}
 \lambda &=& - \epsilon f_0 a^3 \frac{(x_1+x_2)(1+\cos(\psi_1+\psi_2))+l_m(\cos\psi_1+\cos\psi_2)}{A+B \epsilon a^2 } \\
 A&=&(x_1 l_m \sin\psi_1+x_1 x_2 \sin(\psi_1+\psi_2))^2+(x_2 l_m \sin\psi_2+x_1 x_2 \sin(\psi_1+\psi_2))^2\nonumber\\
 B&=&\left(x_1 +l_m \cos\psi_1 + x_2 \cos(\psi_1+\psi_2)\right)^2+\left(x_2 +l_m \cos\psi_2 + x_1 \cos(\psi_1+\psi_2)\right)^2\nonumber
\end{eqnarray}
where the coefficient $A$ vanishes when the filaments are aligned with the motor.
%\begin{equation}
% \lambda = - \epsilon f_0 a \frac{(x_1+x_2)(1+\cos(\psi_1+\psi_2))+l_m(\cos\psi_1+\cos\psi_2)}{(x_1 l_m \sin\psi_1+x_1 x_2 \sin(\psi_1+\psi_2))^2+(x_2 l_m \sin\psi_2+x_1 x_2 \sin(\psi_1+\psi_2))^2} \ .
%\end{equation}
%From Eqs. (\ref{psi1evolutioncaseII}) and (\ref{psi2evolutioncaseII}) with $\epsilon \ll 1$,
%both $\psi_1$ and $\psi_2$ are $O(\epsilon^{1/2})$ or $\pi+O(\epsilon^{1/2})$, or $\gamma=0$.
Several situations can again be distinguished:
\begin{itemize}
\item When the filaments are neither parallel nor antiparallel, 
%the angles $\psi_1$ and $\psi_2$ are related through
%\begin{eqnarray}
%x_1^2+x_2^2+a^2-2 x_1 a \cos\psi_1-2 x_2 a \cos\psi_2+2 x_1 x_2 \cos(\psi_1+\psi_2)=l_m^2
%\end{eqnarray}
%
%Solving for the lagrange multiplier $\lambda$ 
%by imposing the constraint that the length of the motor is equal to $l_m$, 
%and writing the equation in the limit $\epsilon\rightarrow 0$ yields the following dynamic equations
from Eq. (\ref{LambdaEquation}), 
$\lambda\sim \epsilon $ in the limit $\epsilon\rightarrow 0$.
Eqs. (\ref{x1evolutioncaseIII}) and (\ref{x2evolutioncaseIII}) then yield at the lowest order in $\epsilon$
\begin{eqnarray}
 \mu \frac{d x_1}{d t}&=&f_0 + O(\epsilon)\\
  \mu \frac{dx_2}{d t}&=&f_0 + O(\epsilon)
\end{eqnarray}
such that no steady-state exists in that situation.
\item When the filaments are parallel and point in the same direction, $(\psi_1,\psi_2)=(0,\pi)$ or $(\psi_1,\psi_2)=(\pi,0)$. In that case $\lambda=0$ and no steady-state exists for the motor, which runs on the two filaments before detachment.
\item When the filaments are antiparallel, $(\psi_1,\psi_2)=(0,0)$ (expansile configuration) or $(\psi_1,\psi_2)=(\pi,\pi)$ (contractile configuration). 
In the expansile configuration, $x_1+x_2+l_m=\chi a$, with $\chi=\pm1$. 
In the contractile configuration, $x_1+x_2-l_m=\chi a$, with the same rule applying for $\chi$.
In both cases, Eq. (\ref{LambdaEquation}) yields $\lambda=- \chi f_0$ and any position of the motor is a steady-state.

To test for the stability of these steady states, 
we consider the dynamics of $\psi_1$ and $\psi_2$ around $\psi_1=\psi_2=\psi^{*}$ with $\psi^{*} \equiv 0$ or $\pi$,
$\psi_1=\psi^{*}+\delta \psi_1$ and $\psi_2=\psi^{*}+\delta \psi_2$. To regularise the dynamics around the parallel filaments state, we consider a situation with finite $\epsilon$ and consider perturbations verifying $\delta \psi\ll\sqrt{\epsilon}$, such that $\lambda\simeq -\chi f_0$.
From Eqs. (\ref{psi1evolutioncaseIII}) and (\ref{psi2evolutioncaseIII}), we have then
\begin{eqnarray}
 \epsilon \mu a \frac{d\delta \psi_1}{dt}&=&\frac{\chi f_0}{a}
   \left[(x_1 l_m \cos \psi^{*} + x_1 x_2)\delta \psi_1  + x_1 x_2 \delta \psi_2 \right]  \\
 \epsilon \mu a \frac{d\delta \psi_2}{dt}&=&\frac{\chi f_0}{a}
   \left[x_1 x_2 \delta \psi_1 + (x_2 l_m \cos \psi^{*} + x_1 x_2) \delta \psi_2 \right] \ ,
\end{eqnarray}
so that the state $\psi_1=\psi_2=\psi^{*}$ is stable if $\chi ( x_1 l_m \cos \psi^{*} + x_2 l_m \cos \psi^{*} + 2 x_1 x_2) <0$ and $(x_1 l_m \cos \psi^{*} + x_1 x_2)(x_2 l_m \cos \psi^{*} + x_1 x_2)- x_1^2 x_2^2>0$, 
whereas it is unstable otherwise.
Taking the initial condition to be
$x_1=x_{0}/2+(\chi a - l_m \cos \psi^{*})/2$ and $x_2=-x_{0}/2+(\chi a - l_m \cos \psi^{*})/2$ at the steady state, 
the stability condition is satisfied when
\begin{equation}
\chi (x_{0}^2+l_m^2-a^2) > 0 \text{ and } x_1 x_2 \chi \cos \psi^{*} >0 \ .
\end{equation}
%\begin{equation}
% \Big \{ \begin{array}{c}
%           x_1 x_2 >0 \\
%           \chi \cos \psi^{*} > 0
%         \end{array}
%\end{equation} or
%\begin{equation}
%  \Big \{ \begin{array}{c}
%           x_1 x_2 <0 \\
%           \chi \cos \psi^{*} <0
%         \end{array}
%\end{equation}
%for both expansile ($\psi^{*}=0$) and contractile ($\psi^{*}=\pi$) configurations.
%\begin{eqnarray}
% \frac{d\delta \psi_1}{dt}=\frac{4 f_0 l_m \epsilon}{\mu}\left[\frac{-a- x_0 +l_m \epsilon}{(a+x_0 + l_m\epsilon)(a^2+(x_0+l_m\epsilon)^2)\delta \psi_1}+\frac{a+ x_0 +l_m \epsilon}{(a-x_0 + l_m\epsilon)(a^2+(x_0-l_m\epsilon)^2)\delta \psi_2}\right]\\
% \frac{d\delta \psi_2}{dt}=\frac{4 f_0 l_m \epsilon}{\mu}\left[\frac{a-x_0 +l_m \epsilon}{(a+x_0 + l_m\epsilon)(a^2+(x_0+l_m\epsilon)^2)\delta \psi_1}+\frac{-a+ x_0 +l_m \epsilon}{(a-x_0 + l_m\epsilon)(a^2+(x_0+l_m\epsilon)^2)\delta \psi_2}\right]
%\end{eqnarray}
%The expansile configuration is only stable when $x_0<\frac{a}{2}$, that is to say when the center of mass of the motor is located within the two crosslinkers.
%To test for the stability of the contractile configuration, we consider an initial position of the motor $x_1=x_0+\frac{\epsilon l_m-a}{2}$,
%%\begin{eqnarray}
%%\delta \psi_1 \frac{d\delta \psi_1}{dt}=-2 f_0\frac{a^2-4 x_0^2}{a(a+\epsilon l_m+2 x_0)^2}\\
%%\delta \psi_2 \frac{d\delta \psi_2}{dt}=-2 f_0\frac{a^2-4 x_0^2}{a(a+\epsilon l_m-2 x_0)^2}
%%\end{eqnarray}
%The expansile configuration is only stable when $x_0<\frac{a}{2}$, that is to say when the center of mass of the motor is located within the two crosslinkers.
\end{itemize}
The results are summarized in the flow diagram in Fig. \ref{fig:twofilaments}F.
Red segments indicate the stable steady states.
In the green region, there are no possible values of $\psi_1$ and $\psi_2$ for given values of $x_1$ and $x_2$.

\subsection{Two filaments rigidly attached to the external network}
This situation corresponds to case (v).  The mechanical potential $U$ reads as for case (ii)
  \begin{equation}
 U=-f_0(x_1+x_2)+\lambda \left( \sqrt{x_1^2+x_2^2-2 x_1 x_2 \cos\theta}-l_m \right),
 \end{equation}
with $x_1$ and $x_2$ the distances to the crossing point of the two filaments, taken positive in the direction of the filament towards which motors move (Fig. \ref{fig:twofilaments}G). $\theta$ is the angle between the two filaments, 
such that $\mathbf{n}_1.\mathbf{n}_2=\cos\theta$, with $\mathbf{n}_i$ the unit vector giving the orientation of the filament $i$. 
$\lambda$ is a Lagrange multiplier ensuring that the length $l_m$ of the motor is fixed. 
The dynamic equation for the motion of the motor on the two filaments Eq. (\ref{DynamicEquation}) then reads
 \begin{eqnarray}
 \mu \frac{\partial x_1}{\partial t}&=&f_0-\frac{\lambda}{l_m} (x_1-x_2 \cos\theta)\\
  \mu \frac{\partial x_2}{\partial t}&=&f_0-\frac{\lambda}{l_m} (x_2-x_1 \cos\theta),
 \end{eqnarray}
 where $\theta$ is a fixed angle, 
and the Lagrange multiplier $\lambda$ is obtained from the constraint that $x_1^2+x_2^2-2 x_1 x_2 \cos\theta=l_m^2$:
 \begin{equation}
 \lambda=\frac{f_0 l_m (x_1+x_2)(1-\cos\theta)}{(x_1-x_2\cos\theta)^2+(x_2-x_1\cos\theta)^2}
 \end{equation}
 Two solutions can be found for the motor position on the two filament, assuming that they are not parallel ($\theta\neq 0$ and $\theta\neq \pi$):
 \begin{eqnarray}
 x_1=x_2=\pm \frac{l_m}{2\sin\frac{\theta}{2}}
 \end{eqnarray}
 As pointed out in Ref. \cite{Dasanayake11}, a linear stability analysis around the steady-state indicates that only the positive solutions 
$ x_1=x_2= l_m / [2\sin(\theta/2)]$ is stable. The corresponding force dipole at equilibrium is positive and is given by 
\begin{equation}
d=\frac{l_m f_0}{ \sin (\frac{\theta}{2})}.
\end{equation}
% The corresponding average dipole for a uniform distribution of angles $\theta$ is $d=(2l_m f_0)/\pi$.
 Therefore, the dipole formed on two rigidly fixed, non-parallel filaments is always contractile (Fig. \ref{fig:twofilaments}H-I).

\subsection{Averaging force dipole}

\begin{figure}[h]
 \centering
 \includegraphics[width=8cm]{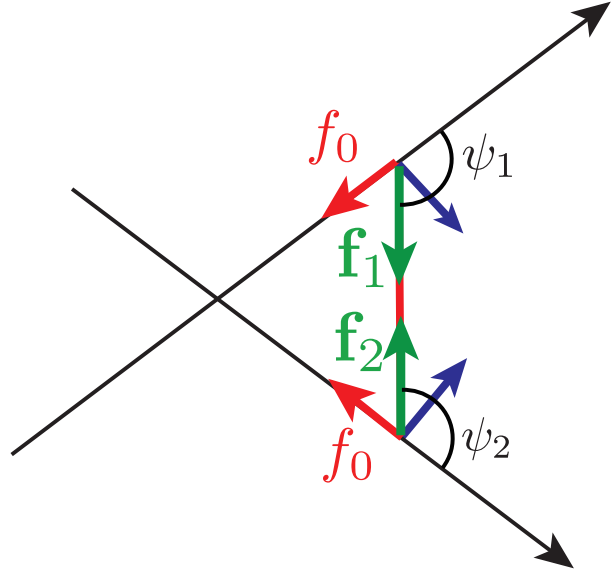}
 \caption{Schematic of the force dipole exerted by a motor. Each motor exerts a force of magnitude $f_0$ parallel to the filament. The total force exerted by the motor on the filament are equal and opposite at steady-state and denoted $f_1$ and $f_2$. The contribution of the force normal to the filament arises from geometrical constrains. Overall, this results in a tension $f=|\mathbf{f}_1|=|\mathbf{f}_2|$ acting within the motor, which is used here to define the force dipole exerted by a motor.
 \label{fig:forcedipoleschematic}}
\end{figure}

To obtain the associated average force dipole exerted by the motor in these different configurations, we proceed as follows: 
denoting as $\mathbf{c}_{12}^{(0)}$ the initial distance between the two filaments, 
$\theta_{f,1}^{(0)}$ and $\theta_{f,2}^{(0)}$ the two angles giving the initial orientations of the filaments, 
$x_{f,1}^{(0)}$, $x_{f,2}^{(0)}$ the initial position of the two motors on the filaments,
and $r_{k}$ ($k=1,\cdots,N_c$ with $N_c$ the number  of crosslinkers attached to the external network) 
the position of the crosslinkers on the filaments,
we compute the motor-induced force dipole $d$ reached at steady-state as shown in Fig. \ref{fig:twofilaments}; 
cases (ii), (iii) and (v) by 
\begin{equation}
 d(\mathbf{c}_{12}^{(0)},\theta_{f,1}^{(0)},\theta_{f,2}^{(0)},x_{1}^{(0)},x_{2}^{(0)},\{r_{k} \}_k)
  = \frac{l_m f_0 }{2 \cos \psi_1} + \frac{l_m f_0 }{2 \cos \psi_2} \ , \label{eq:forcedipoledefinition}
\end{equation}
%which is also a function of  $x_{f,1}^{(0)}$, $x_{f,2}^{(0)}$.
where the angles $\psi_1$ and $\psi_2$ between the motor and two filaments at steady-state
are functions of $\mathbf{c}_{12}^{(0)}$, $\theta_{f,1}^{(0)}$, $\theta_{f,2}^{(0)}$, $x_{1}^{(0)}$, $x_{2}^{(0)}$ and $\{r_{k} \}_k$.
%If a steady-state is not reached because the motor is detaching from the filament, the corresponding force dipole is counted as a $0$ contribution.
Equation (\ref{eq:forcedipoledefinition}) is obtained from the definition of the force dipole
\begin{equation}
d = -\frac{l_m}{2} \mathbf{n} \cdot (\mathbf{f}_1 - \mathbf{f}_2)
\end{equation}
with $\mathbf{n}$ the unit vector giving the motor orientation pointing towards filament 1,  and $\mathbf{f}_i$ with $i=1,2$ are the forces exerted by the motor on the two filaments.
The dependencies $1/\cos \psi_i$ ($i=1,2$) in Eq. (\ref{eq:forcedipoledefinition}) arise from 
the condition that the projections of the forces $\mathbf{f}_i$ on the filament $i$ have magnitude $f_0$ (Fig. \ref{fig:forcedipoleschematic}).

The average force dipole is then obtained by
\begin{equation}
<d>=\int d \mathbf{c}_{12}^{(0)} 
\int_0^{2\pi} \frac{d \theta_{f,2}^{(0)}}{2\pi} \int_0^{2\pi} \frac{d \theta_{f,2}^{(0)}}{2\pi} 
\int \left( \prod_{k=1}^{N_c} \frac{d r_{k}}{L_f} \right) \int \frac{d (x_{1}^{(0)}, x_{2}^{(0)})}{\Gamma}
d(\mathbf{c}_{12}^{(0)},\theta_{f,1}^{(0)},\theta_{f,2}^{(0)},x_{1}^{(0)},x_{2}^{(0)},\{r_{k} \}_k) \ ,
 \end{equation}
where the integration $\int d (x_{1}^{(0)}, x_{2}^{(0)})$ runs 
for all possible initial motor configurations $x_{1}^{(0)}$, $x_{2}^{(0)}$
that eventually reach a steady state,
and $\Gamma = \int d (x_{1}^{(0)}, x_{2}^{(0)}) 1$.

\section{Relationship between the stress and the average force dipole} \label{StressAverageForceDipole}

In this section, we discuss the two contributions to the stress generated by the filament and motor network, $\sigma_{ij}^f$ and $\sigma_{ij}^m$. We point out that in the absence of large scale deformations, $\sigma_{ij}^f$ can contribute to the total stress, when the elastic response of the filament network is non-linear. 
\begin{figure}[h]
 \centering
 \includegraphics[width=14cm]{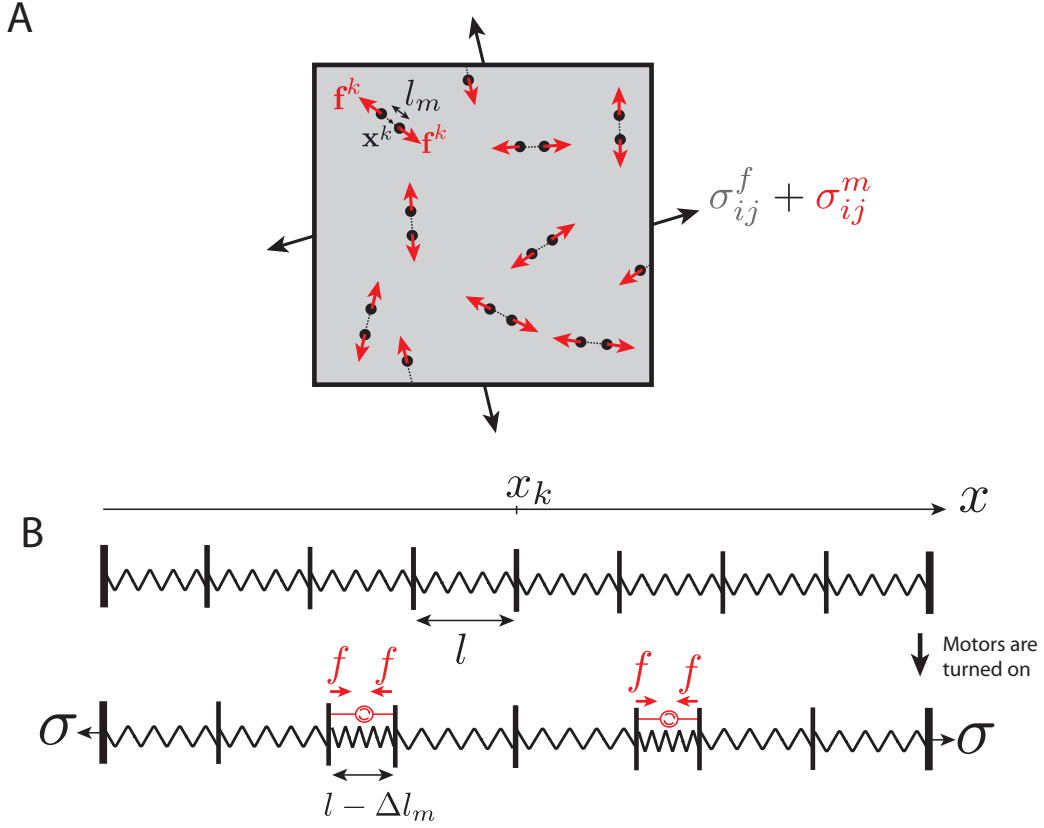}
 \caption{A. Schematic of a 2D elastic material, subjected to forces arising from force dipoles exerted by motor filaments. Motors result in a stress $\sigma_{ij}$ acting at the boundary of the box. 
 B. Schematic of a 1D chain of non linear springs subjected to force dipoles. The chain contains $N$ springs and is fixed at both ends. Motors result in a force $\sigma$ along the chain.
 \label{fig:1Dchain}}
\end{figure}

\subsection{Two-dimensional, linear elastic material}

We consider here a two-dimensional elastic material subjected to the force dipoles exerted by motor filaments. The motor force dipoles consist of two opposite forces $\mathbf{f}^k=\pm f^k\mathbf{n}^k$, separated by a distance $l_m$, and where the index $k$ label the motors. 
The motors induce a deformation in the elastic material. The total stress in the system is then the sum of the resulting elastic stress, denoted $\sigma_{ij}^f$, and the stress generated within the motors, denoted $\sigma_{ij}^m$:
\begin{equation}
\sigma_{ij}=\sigma_{ij}^f+\sigma_{ij}^m
\end{equation}

We start by discussing the average stress created by an ensemble of motors in a 2D material. Following Ref. \cite{aditi2002hydrodynamic}, we now show that when the system is homogeneous, force balance on a section $\mathcal{S}$ of the network enclosed in a contour $\mathcal{C}$ allows to relate the stress generated within motors to the concentration and orientation of motors. To evaluate the stress, we first note that each motor $k$ corresponds to a line under tension $f^k$ with length $l_m$, orientation $\mathbf{n}^k$, centered at position $\mathbf{x}^k$, and joining the two points at coordinates $\mathbf{x}^k-l_m/2 \mathbf{n}^k$ and $\mathbf{x}^k+l_m/2 \mathbf{n}^k$  (Fig. \ref{fig:1Dchain}A). 
Therefore, the two-dimensional stress field within the motors can be written
\begin{eqnarray}
\sigma_{ij}^m(\mathbf{x})=\sum_k f^k n^k_i n^k_j\int_{-\frac{l_m}{2}}^{\frac{l_m}{2}} du \delta(\mathbf{x}-(\mathbf{x}^k+\mathbf{n}^k u))
\end{eqnarray}
where $u$ is a coordinate going along the line under tension $f^k$.
The average stress within a region $\mathcal{S}$ of surface area $S$ is then given by
\begin{eqnarray}
\langle \sigma_{ij}^m \rangle&=&\frac{1}{S}\int_{\mathcal{S}} d\mathbf{x} \sigma^m_{ij}(\mathbf{x})\\
&=&\frac{1}{S} \int_{\mathcal{S}} d\mathbf{x} \sum_k f^k n^k_i n^k_j\int_{-\frac{l_m}{2}}^{\frac{l_m}{2}} du \delta(\mathbf{x}-(\mathbf{x}^k+\mathbf{n}^k u))\\
&=&\frac{1}{S} \sum_k f^k l_m n_{i}^k  n_{j}^k \\
&=&c \langle  d  n_{i}  n_{j} \rangle \label{EqStressForceDipole}
\end{eqnarray}
where $c$ is the concentration of motors, $d^k=l_m f^k$ is the dipole strength of motor $k$, and the averaging $\langle\cdot\rangle$ is performed over space.
%When considering the stress generated in a box of size $L$, the stress measured on the boundary of the box approaches $\langle \sigma_{ij}^m\rangle$ as $L\rightarrow \infty$. Therefore, $\langle \sigma_{ij}^m\rangle$ gives the coarse-grained stress generated within the motors.

%\begin{eqnarray}
%\int_C dl_\alpha  \sigma^m_{\alpha\beta}&=&\int_S d\mathbf{x} \sum_k \frac{d^k}{l_m} n_{\beta}^k\nonumber\\
%&& \left[\delta(\mathbf{x}-(\mathbf{x}^k+\mathbf{n}^k l_m))-\delta(\mathbf{x}-(\mathbf{x}^k-\mathbf{n}^k l_m))\right]\\
%\int_S d\mathbf{x} \partial_{\alpha}\sigma^m_{\alpha\beta}&\simeq&- \int_S d\mathbf{x}\sum_k d^k  n_{\beta}^k n_{\gamma}^k\partial_{\gamma} \delta(\mathbf{x}-\mathbf{x}_k) \\
%\int_S d\mathbf{x} \partial_{\alpha}\sigma^m_{\alpha\beta}&\simeq& \int_S d\mathbf{x}\partial_{\gamma} (\sum_k  d^k  n_{\beta}^k n_{\gamma}^k)\delta(\mathbf{x}-\mathbf{x_k}) \\
%\sigma_{\alpha\beta}&\simeq& <d   n_{\alpha}n_{\beta} c>\label{EqStressForceDipole}
%\end{eqnarray}

In a linearly elastic material, the average stress within the network of filaments is given by 
\begin{equation}
\sigma_{ij}^f=2E \left( u_{ij}-\frac{1}{2}u_{kk}\delta_{ij}\right)+K u_{kk} \delta_{ij} 
\end{equation}
with $E$ and $K$ a shear and bulk elastic moduli, and $u_{ij}=\frac{1}{2}(\partial_{i}u_{j}+\partial_{j}u_{i})$ the gradient of deformation. The average stress generated in the elastic material in a region $\mathcal{S}$ with contour $\mathcal{C}$ and surface area $S$  is then given by
\begin{eqnarray}
\langle\sigma_{ij}^f \rangle&=&\frac{1}{S}\int_{\mathcal{S}} d\mathbf{x} \sigma^f_{ij}\\
&=&\frac{2E}{S} \int_{\mathcal{S}} d\mathbf{x} u_{ij}+\frac{K-E}{S} \int_{\mathcal{S}} d\mathbf{x} u_{kk} \delta _{ij}\\
&=&\frac{E}{S}\int_{\mathcal{C}} dl \nu_{i} u_{j}+\frac{E}{S}\int_{\mathcal{C}} dl \nu_{j} u_{i}+\frac{K-E}{S}\int_{\mathcal{C}} dl \nu_k u_k \delta_{ij}\label{AverageFilamentStress}
\end{eqnarray}
where $\boldsymbol{\nu}$ is the vector normal to the contour $\mathcal{C}$ and $dl$ an infinitesimal line element on the contour. If we consider a square box whose boundaries are fixed (Fig. \ref{fig:1Dchain}A), or periodic boundary conditions, the contour integrals in Eq. (\ref{AverageFilamentStress}) vanish. Therefore, for a linear elastic material with fixed boundaries, the average stress arises entirely from the forces acting within the motors. This however does not apply to a non-linearly elastic material, as we show in the next section.

\subsection{One-dimensional chain of non-linear springs}

We consider a simpler example in 1D of a periodic chain of $N$ elastic springs. Each spring is located between positions $x_{i}$ and $x_{i+1}$, with initial resting position $x_i=il$. The two points at the end of the chain, $x_0$ and $x_N$, are not allowed to move. In addition, $N_m$ motors are acting in parallel to a fraction $n=N_m/N$ of the springs (Fig. \ref{fig:1Dchain}B). The motor exerts a constant force $f$. The springs have a non-linear force-extension relation
\begin{equation}
f_e=k \frac{\Delta l}{l}+ k_2 \left(\frac{\Delta l}{l}\right)^2
\end{equation}
with $f_e$ the force exerted by the spring, $\Delta l=x_{i+1}-x_i-l$ the extension of the spring, and $k$ and $k_2$ are two spring constants. We assume that the springs are weakly non-linear, $k_2/k \ll 1$.
In the initial resting position, $\Delta l=0$. The motors are then turned-on, driving a deformation of the the springs in the chain.  The contraction of the springs which are in parallel with the motors is denoted $-\Delta l_m$. Because the overall length of the chain is kept fixed, the deformation of the free springs is then given by $\Delta l_m n/(1-n)$. We denote by $\sigma$ the total force acting within the chain. Force balance imposes that the total force within the springs and the motors is fixed and equal to $\sigma$, giving
\begin{eqnarray}
\sigma&=& - k \frac{\Delta l_m}{l} + k_2 \left(\frac{\Delta l_m}{l}\right)^2 +f=k \frac{\Delta l_m}{l}\frac{ n}{1-n}+k_2 \left(\frac{\Delta l_m}{l}\frac{ n}{1-n}\right)^2
\end{eqnarray}
where the second part of the equality is the total force within the springs in parallel with a motor, and the third part the force within the free springs.
Solving these equations and expanding to first order in $f k_2/k^2 \ll 1$, one obtains
\begin{eqnarray}
\sigma=f n+\frac{k_2 f^2 n(1-n)}{k^2} 
\end{eqnarray}
or using the concentration of motors $c=n/l$,
\begin{eqnarray}
\sigma&=&f l c+\frac{k_2 f^2  }{k^2 } cl(1-cl)\\
&=&\sigma^m+\sigma^f
\end{eqnarray}
where $\sigma^m=flc$ is the average tension exerted within the motors. To leading order in the spring non linearity, $k_2\rightarrow 0$, the force within the chain reduces to the average motor tension $\sigma^m$, in agreement with Eq. (\ref{EqStressForceDipole}). The non-linear elastic behaviour however brings a correction to this term $\sigma^f=k_2 f^2 cl(1-cl)/k^2$, proportional to the motor force squared, $f^2$. 
%This additional contribution is positive when $k_2>0$, i.e. when the springs are strain-stiffening \cite{liverpool2009mechanical, e2011active}. Therefore, the stress exerted by motors in a non-linear elastic material does not reduce in general to $\sigma^m$.

\section{Network relaxation time} \label{App:FirstPassageTime}

\begin{figure}[h]
 \centering
 \includegraphics[width=12cm]{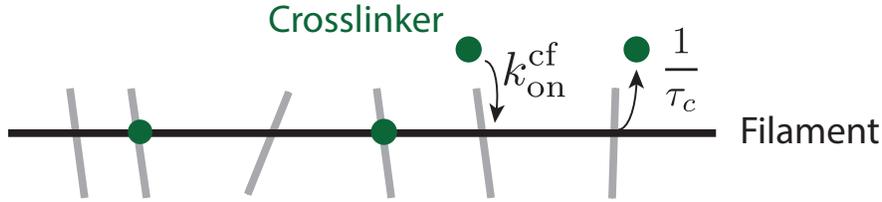}
 \caption{Schematic of a filament in the network. Crosslinkers bind  to the filament with rate $k_{on}^{\rm{cf}}$ and unbind with rate $1/\tau_c$.
 \label{fig:1Dchain_crosslinker}}
\end{figure}

We derive here an approximate expression for the network characteristic relaxation time. We consider a filament within the network, crossing other  filaments that are themselves immobilised. We expect this last assumption to be valid for a large enough number of crosslinkers. The number of crossing points is assumed to largely exceed the total number of crosslinkers in the network. Crosslinkers bind and unbind the filament at crossing points with other filaments. A filament with two attached crosslinkers is completely fixed, and can only rotate when it has one attached crosslinker. For simplicity we consider here that the filament can not rearrange to relax stresses unless no crosslinker attaches it to other filaments in the network. Once a filament is free, it can move until a crosslinker binds to it and immobilise it. We therefore estimate the network relaxation time as the mean first passage time to a state where no crosslinkers bind the filament, from a state where one crosslinker binds the filament.

We consider the probability of having $n_c$ crosslinkers on the filament, $P(n_c)$. 
New crosslinkers bind to the filament with rate 
$k_{on}^{\rm{cf}}=2 [k_{on}^{\rm{c}}/(1+k_{on}^{\rm{c}}\tau_c)]  N_c/N_f$, 
as crosslinkers bind to all filaments in the network with rate $k_{on}^{\rm{c}}$ (Fig. \ref{fig:1Dchain_crosslinker}), and each crosslinker binds two filaments. In addition, crosslinkers unbind from the filament with rate $\tau_c^{-1}$ (Fig. \ref{fig:1Dchain_crosslinker}). The probability $P(n_c)$ then follows the master equation:
\begin{align} 
\frac{d P(n_c)}{dt} &=  \frac{n_c+1}{\tau_c} P(n_c+1) + k_{on}^{\rm{cf}} P(n_c-1) - \left(\frac{n_c}{ \tau_c}+k_{on}^{\rm{cf}} \right) P(n_c)
\label{eq:supp_masterequation}
\end{align}

\paragraph{Stationary distribution.}
The stationary distribution of Eq. (\ref{eq:supp_masterequation}) is a Poisson distribution:
%\begin{equation}
%P(n_c)=P_0\prod_{k=1}^{n_c} \frac{k_{on}^c\tau_c}{k}=P_0\frac{(k_{on}^c\tau_c)^{n_c}}{n_c!}
%\end{equation}
%$P_0$ is obtained from the condition $\sum_{n_c} P(n_c)=1$, so that $P_0=e^{-k_{on}^c\tau_c}$ and
\begin{equation}
P(n_c)=e^{-k_{on}^{\rm{cf}}\tau_c}\frac{(k_{on}^{\rm{cf}}\tau_c)^{n_c}}{n_c!}
\end{equation}
%which is the Poisson distribution, 
%with mean and variance $n_c^*=k_{on}^{\rm{cf}}\tau_c=2k_{on}^{\rm{c}} \tau_c/N_f$.
with mean and variance $n_c^*=k_{on}^{\rm{cf}}\tau_c \simeq 2N_c/N_f$ for $k_{on}^{\rm{c}}\tau_c \gg 1$.

\paragraph{Mean first passage time.}

We now want to obtain the mean first passage time to reach a state where the filament has no bound crosslinker, $n_c=0$, starting from a configuration with $n$ crosslinkers attached on the filament.  We denote $T_n$ this first passage time, and we follow a standard procedure to obtain its value \cite{gardiner1985stochastic, pury2003mean}. $T_n$ satisfies the following equation for $n>0$:
\begin{eqnarray}
\label{MasterEquationFirstPassageTime}
T_n=\frac{k_{on}^{\rm{cf}}}{k_{on}^{\rm{cf}}+n/\tau_c}T_{n+1}+\frac{n/\tau_c}{k_{on}^{\rm{cf}}+n/\tau_c} T_{n-1}+\frac{1}{k_{on}^{\rm{cf}}+n/\tau_c}
\end{eqnarray}
where the first term corresponds to the probability of moving to a state with $n+1$ bound crosslinkers, times the waiting time from the state $n+1$, the second term is the product of the probability to move to a state with $n-1$ bound crosslinkers times the waiting time from the state $n-1$, and the last term is the average time spent in the state $n$. In addition, $T_0=0$ by definition of the first passage time $T_n$, and 
we take a reflecting boundary condition at infinity, implying $T_n-T_{n-1}\rightarrow 0$ for $n\rightarrow \infty$.
Equation (\ref{MasterEquationFirstPassageTime}) can be rewritten
\begin{eqnarray}
k_{on}^{\rm{cf}}(T_n-T_{n+1})+\frac{n}{\tau_c} (T_n-T_{n-1})=1
\end{eqnarray}
or defining $z_n=T_{n+1}-T_n$,
\begin{eqnarray}
z_n=\frac{n}{k_{on}^{\rm{cf}} \tau_c} z_{n-1}-\frac{1}{k_{on}^{\rm{cf}}}
\end{eqnarray}
To solve this equation, one introduces $Z_n=z_n (k_{on}^{\rm{cf}}\tau_c)^n / n! $, which satisfies then:
\begin{equation}
\label{EqZn}
Z_n=Z_{n-1}- \frac{ (k_{on}^{\rm{cf}}\tau_c)^n}{k_{on}^{\rm{cf}}n!}
\end{equation}
Solving Eq. (\ref{EqZn}) then yields the following expression for $Z_n$, $z_n$ and $T_n$, using that $Z_{\infty}=0$: %(assuming here $Z_0=0$ based on $T_{-1}=0$)
\begin{eqnarray}
Z_n&=&\sum_{k={n+1}}^{\infty} \frac{ (k_{on}^{\rm{cf}}\tau_c)^k}{k_{on}^{\rm{cf}}k!}\\
z_n&=&\frac{n!}{(k_{on}^{\rm{cf}}\tau_c)^n}\sum_{k={n+1}}^{\infty} \frac{ (k_{on}^{\rm{cf}}\tau_c)^k}{k_{on}^{\rm{cf}}k!}\\
T_n&=&\sum_{k=0}^{n-1} \frac{k!}{(k_{on}^{\rm{cf}} \tau_c)^k}\sum_{m={k+1}}^{\infty} \frac{ (k_{on}^{\rm{cf}}\tau_c)^m}{k_{on}^{\rm{cf}} m!}
\end{eqnarray}
From this last expression, one finally obtains $T_1$, 
\begin{eqnarray}
T_1&=&  \sum_{m=1}^{\infty} \frac{ (k_{on}^{\rm{cf}} \tau_c)^m}{k_{on}^{\rm{cf}} m!}\\
&=&\frac{1}{k_{on}^{\rm{cf}}} (e^{k_{on}^{\rm{cf}} \tau_c }-1)\\
&=&\frac{\tau_c}{n_c^*} (e^{n_c^* }-1)
\end{eqnarray}

The time $T_1$ therefore contains a factor increasing exponentially with the number of crosslinkers per filament $n_c^*$. We use the time $T_1$ as an estimate for the network relaxation time. In Fig. 3f, we verify that the time $T_1$ provides a good approximation for the relaxation time of the network.

\section{Effect of motor concentration and dispersion in motor forces}

\begin{figure}[h!]
 \centering
 \includegraphics[width=14cm]{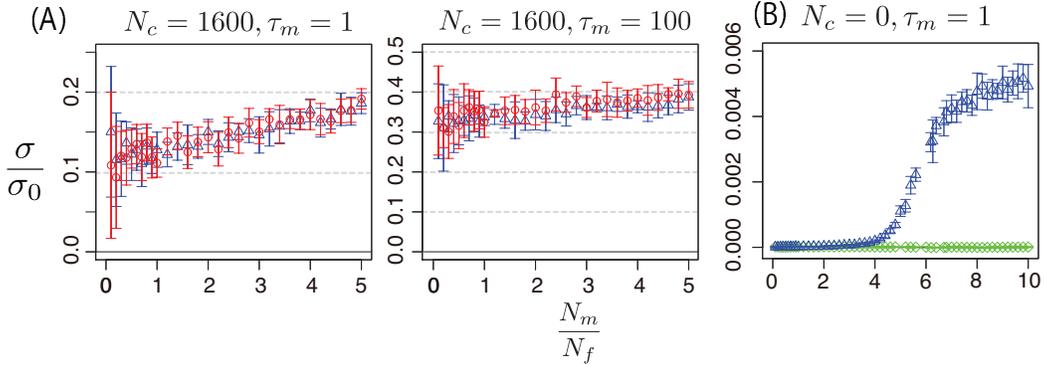}
 \caption{A. Isotropic stress as a function of the number of motors $N_m$, for two values of myosin turn-over. 
In these graphs, we performed the simulations for the cases with identical motors (red circles) and
with the a dispersion of 20\% motor force and friction (blue triangles).
Non-specified parameters are as in Fig. 3 of the main text. 
B. Isotropic stress as a function of the number of motors, in the absence of a cross linker. 
Green diamond marks, identical motors; no stress is generated. Blue triangle marks, a dispersion of 20\% motor force and friction results in a non-zero stress for a high enough number of motors. Non-specified parameters are as in Fig. 3 of the main text.
Stress was averaged here over $10\tau$ after initialization.
 \label{fig:myosindependency}}
\end{figure}

We have performed additional simulations measuring the isotropic stress $\sigma$ as the concentration of myosin motors $N_m/W^2$ is varied (Figure. \ref{fig:myosindependency}A). The reference stress $\sigma_0$ is defined as being proportional to motor concentration, such that $\sigma\sim\sigma_0$ indicates that the stress is proportional to the myosin concentration. We find that $\sigma$ has a weak non-linear dependence as a function of the myosin number $N_m$ for the simulation parameters plotted in Figure \ref{fig:myosindependency}A, and is nearly linearly increasing with $N_m$ for slow enough turn-over. This indicates that the stress generated by the simulated network is roughly proportional to the myosin concentration.

In our simulations, motors are identical and have the same stall force; as a result no stress is generated in the absence of cross linkers (Fig. 3 and Ref \cite{lenz2012requirements}). Introducing a dispersion in motor friction and stall force however can result in stress generation as the number of myosin motors is increased (Figure. \ref{fig:myosindependency}B). The overall stress generated is much smaller than stresses generated for the same parameters with a cross linked network.

\section{Details of the model used for the numerical simulation}

\subsection{Main Components}

We introduce three components in our simulations: actin filaments, crosslinkers, and myosin filaments. 

Myosin filaments are represented by rigid rods with a length $l_m$. To represent the attachment between myosin and actin filaments, two springs are added at the end of each myosin mini filament with stiffness $k_{ms}$ and reference length $0$, such that the tension in the end spring $f_{ms}$ is $f_{ms}=k_{ms} l_{ms}$, with $l_{ms}$ the length of the connecting spring. In practice $k_{ms}$ is taken to be large enough that the spring maximum extension is very small compared to other lengths in the simulation. The position of a myosin mini filament $k$ on a filament $i$, relative to the center of mass of the filament $i$, is denoted $s_{ki}$.
  
Actin filaments are treated as rigid rods with a finite length, $l_f$.  The position of the centre of mass of the filament is denoted $\mathbf{x}_{f,i}$ and the  unit vector giving the orientation of filament $i$ is denoted $\mathbf{n}_{f,i}$.

Crosslinkers are represented by springs connecting two actin filaments, with a stiffness $k_x$ and reference length $0$, such that the tension with an crosslinker is $f_x=k_x l_x$ with $l_x$ the length of the crosslinker. In practice $k_x$ is chosen large enough such that the length of the cross linker is very small compared to other lengths in the simulation.

%The simulation is a 3D simulation of a thin polymer network with periodic boundary
%conditions in the x and y directions, and free boundaries the z-direction.
We simulate the network in two dimensions $x,y$, in a periodic box of size $W$. Simulations are initialised by positioning filaments in the box with random positions and orientations.

\subsection{Dynamics}

In the simulations, bond myosin minifilaments move on actin filaments. 
Myosin filaments detach from a filament when they reach the end of a filament. In addition, when myosin turnover is taken into account, they can spontaneously bind and unbind filaments. Crosslinkers can also bind and unbind filaments when crosslinker turnover is taken into account, and filaments can be removed and added in the network when filament turnover is taken into account. At every step of myosin motion, the network is relaxed quasi-statically to equilibrium. 

\subsubsection{Myosin motion}
%The simulation is broken into two phases, a slow kinetics phase and a fast dynamics phase.
%Actin, crosslinkers and myosin minifilaments turnover and update during the slow kinetics phase. 
%During the dynamics phase the simulation is relaxed to static 
%equilibrium. The dynamics are fast compared to the kinetics.

At every step, the end position of the myosin minifilament $k$ on filament $i$, denoted $s_{ki}$, is updated according to the following equation:
\begin{equation}
\mu \frac{ds_{ki}}{dt} = f_0 -\mathbf{f}_{ks} \cdot \mathbf{n}_i.
\end{equation}
where $\mathbf{f}_{ks}$ is the tension of the myosin end-spring connected to the filament,  $f_0$ is the active force generated by the myosin on the filament, and $\mu$ is an effective friction coefficient between the motor and the filament. The equation above is discretised with an Euler explicit scheme with time step $dt$.
 
%If $|s|>l_m/2$ then the myosin has walked off of the end of the filament and the head becomes
%un-attached.

\subsubsection{Filament and crosslinker turnover}

Crosslinkers are added with a rate $k_{on}^c$ and removed with a rate $k_{off}^c=1/\tau_c$. The number of bound cross linkers $N_{cb}$ and unbound cross linkers $N_{cu}$ add up to the total number of cross linkers $N_c=N_{cb}+N_{cu}$. At every step, each unbound cross linker has a probability of being added to the network $k_{on}^c dt$, and each bound cross linkers has a probability to be removed from the network $k_{off}^c dt$. Unbound cross linkers are added by looking randomly for a free cross linking point on two filaments, and attaching the crosslinker there.

Similar rules apply to turnover of myosin and actin filaments.
The binding positions of two ends of the myosin filament are determined in the following way: 
Firstly, an actin filament and the position on it are randomly chosen, and one end of the myosin filament attached there.
Then, the position of the second end of myosin filament is randomly chosen from all the attacheable positions,
{\it i.e.} all the positions on actin filaments located $l_m$ away from the position on which the other end is attaching.

\subsubsection{Quasistatic relaxation}

At every time step, the network is relaxed quasi-statically to equilibrium.  Below, we denote $t^*$ a fictitious time coordinate used for quasi-static relaxation. Quasi-static relaxation is performed by updating the filament center of mass $\mathbf{r}_i$ and orientation $\mathbf{n}_i=(\cos\theta_i,\sin\theta_i)$ according to the following equation:

\begin{eqnarray}
  \alpha_t \frac{d\mathbf{r}_i}{d t^*} & = & \sum_k \mathbf{f}_{ik} \label{TranslationQuasiStatic}\\
  \alpha_r \frac{d\theta_i}{dt^*} & = & \mathbf{e}_z\cdot\left(\sum_k s_k \mathbf{n}_i\times \mathbf{f}_{ik}\right)\label{RotationQuasiStatic}
\end{eqnarray} 
where $\mathbf{f}_{ik}$ is the force acting on filament $i$ from the crosslinker or motor $k$. $\alpha_t$ and $\alpha_r$ are two translational and rotational fictitious friction coefficients, used for the quasi-static relaxation, and $\mathbf{e}_z$ is the direction orthogonal to the plane of simulation. The same equations are used to iterate the position of myosin motors, $\mathbf{r}_k$ and orientation $\mathbf{n}_k$.

Iteration is performed by discretising equations (\ref{TranslationQuasiStatic}) and (\ref{RotationQuasiStatic}) with an Euler explicit scheme, until the system reaches quasi-static equilibrium. In practice, a criterion must be used to specify when the system is close enough to equilibrium. To do so, the squares of the filament translational and rotational velocities are averaged according to:
\begin{equation}
e = \frac{1}{N}\left[
\sum_i \left(\frac{d\mathbf{r}_i}{d t^*}\right)^2+ \frac{l_f^2}{12} \sum_i  \left(\frac{d\theta_i}{d t^*}\right)^2
+\sum_k \left(\frac{d\mathbf{r}_k}{d t^*}\right)^2+ \frac{l_m^2}{12} \sum_k  \left(\frac{d\theta_k}{d t^*}\right)^2
\right]
\end{equation}
where $N$ is the total number of actin and myosin filaments. 
The quasi-static relaxation is completed when $e$ is less than a threshold parameter $e_{max}$.

%Acknowledgements
%.........(Acknowledgements)........

\section{Supplementary movie legends}
\begin{itemize}
\item {\it Supp. Movie M1} Simulation of a network with no crosslinker turnover, no filament turnover, $N_c=800$ and $\tau_m=100\tau$. Total simulation time, $1600\tau$. The number of cross linkers $N_c$ is below the threshold for the network to exert a contractile stress.
\item  {\it Supp. Movie M2} Simulation of a network with no crosslinker turnover, no filament turnover, $N_c=1100$ and $\tau_m=100\tau$. Total simulation time, $1600\tau$. The number of cross linkers $N_c$ is above the threshold for the network to exert a contractile stress.
\item {\it Supp. Movie M3} Simulation of a network with no filament turnover,  crosslinker turnover with $\tau_c=100\tau$, $N_c=1200$ and $\tau_m=100\tau$. Total simulation time, $1600\tau$. The network collapses and does not exert a contractile stress in steady state.
\item {\it Supp. Movie M4} Simulation of a network with filament turnover with $\tau_a=100\tau$, crosslinker turnover with $\tau_c=100\tau$, $N_c=800$ and $\tau_m=100\tau$. Total simulation time, $1600\tau$. The network reaches a steady-state where no contractile stress is exerted.
\item {\it Supp. Movie M5}  Simulation of a network with filament turnover with $\tau_a=100\tau$, crosslinker turnover with $\tau_c=100\tau$, $N_c=1200$ and $\tau_m=100\tau$. Total simulation time, $1600\tau$. The network reaches a steady-state where a contractile stress is exerted.
\end{itemize}

\bibliographystyle{unsrt}
\bibliography{HiraiwaSalbreux_Arxiv}

\end{document}